\documentclass[a4paper,11pt]{JHEP3}
\usepackage{fancyheadings}
\usepackage[dvips]{graphicx}
\usepackage{ifthen}
\usepackage{amsmath}
\usepackage{epsfig}
\usepackage[errorshow]{tracefnt}
\renewcommand{\theenumii}{\alph{enumii}}

\newcommand\nn{\nonumber}

\newcommand\be{\begin{equation}}
\newcommand\beal{\begin{align}}
\newcommand\eeal{\end{align}}
\newcommand\benu{\begin{enumerate}}
\newcommand\eenu{\end{enumerate}}
\newcommand\bit{\begin{itemize}}
\newcommand\eit{\end{itemize}}

\newcommand{\slsh}[1]{\displaystyle{\not} #1}

\newcommand{\ee}{\end{equation}}
\newcommand{\bd}{\begin{displaymath}}
\newcommand{\ed}{\end{displaymath}}

\newcommand{\la}{\lambda}
\newcommand\al{\alpha}
\newcommand\bn{\beta}

\newcommand\ga{\gamma}
\newcommand\de{\delta}

\newcommand\e{\epsilon}

\newcommand{\upa}{\! \begin{array}{c} \alpha\\{} \end{array}}
\newcommand{\upb}{\!\!\begin{array}{c} \beta\\{} \end{array}}
\newcommand{\upc}{\!\!\begin{array}{c} \gamma\\{} \end{array}}
\newcommand{\upd}{\!\!\begin{array}{c} \delta\\{} \end{array}}

\def\ub{\underline{\phantom{\alpha}}\!\!\!\beta}
\def\uc{\underline{\phantom{\alpha}}\!\!\!\gamma}
\newcommand{\una}{\underline{\alpha}}

\newcommand{\und}{\underline{\delta}}
\newcommand{\une}{\underline{\epsilon}}

\title{Massive IIA supergravities}

\author{Dimitrios Tsimpis \\
Max-Planck-Institut f\"{u}r Physik --Theorie\\
F\"{o}hringer Ring 6,  80805 M\"{u}nchen, Germany\\
E-mail: \email{tsimpis@mppmu.mpg.de}
}

\abstract{We perform a systematic search for 
all possible massive deformations of IIA 
supergravity in ten dimensions. 
We show that there exist exactly two possibilities: Romans supergravity and 
Howe-Lambert-West supergravity. Along the way we give the full 
details of the ten-dimensional superspace formulation of the latter. 
The scalar superfield at canonical mass dimension zero 
(whose lowest component is the dilaton), present in
both Romans and massless IIA supergravities, 
is  not introduced
from the outset but 
its existence follows from a certain integrability condition
implied by the Bianchi identities. This fact leads to the 
possibility for a certain topological modification of 
massless IIA, reflecting an analogous situation in eleven dimensions. }

\preprint{MPP-2005-87}
\begin{document}

\section{Introduction}

Romans massive supergravity \cite{roma} has attracted a lot of interest 
following the observation that its mass parameter (cosmological constant) 
may be thought of as sourced by  the D8 brane of type IIA \cite{polc}. If string theory 
is to be understood as embedded in M-theory, it would be desirable 
to have an eleven-dimensional understanding of Romans supergravity. 
The latter, however, has no covariant eleven-dimensional lift\footnote{See \cite{blo} for a 
noncovariant embedding of Romans supergravity in 
eleven-dimensions, and \cite{hs} for a recent implementation 
of the same idea in eleven-dimensional superspace. In \cite{hull} 
it was argued that although Romans massive IIA supergravity 
cannot be embedded in ordinary eleven-dimensional supergravity, 
massive IIA {\it string theory} can be embedded in M-theory.}, 
owing to certain no-go theorems 
forbidding any straightforward introduction of mass in 
eleven-dimensions \cite{dese}.

There does exist however a topological modification 
of eleven-dimensional supergravity  
(subsequently dubbed `MM-theory'),  
as pointed out by Howe in \cite{h},  
which allows the introduction of 
a mass parameter upon compactification to ten dimensions. 
In this way one obtains   
the Howe-Lambert-West supergravity of reference \cite{hlw}. 
The latter  contains a lot of interesting physics \cite{cpr, cla, clb}; it is 
nevertheless much less studied despite  
that, contrary to Romans supergravity, it has a 
well-understood covariant eleven-dimensional origin and  
it has been shown to admit a de Sitter vacuum\footnote{  
Presumably this apparent neglect should 
be attributed to certain unconventional features of HLW supergravity. For example,  
its equations-of-motion cannot be integrated to 
a local Lagrangian. More importantly, it is not clear at present 
if the theory can be made quantum-mechanically self-consistent.}. 
HLW supergravity can also be obtained by a generalized Scherk-Schwarz 
reduction of the equations-of-motion of 
ordinary eleven-dimensional supergravity \cite{llp}.

It would be desirable to have an understanding of the relation between 
Romans and HLW supergravity from a purely ten-dimensional perspective. 
We would also like to know how unique these supergravities are and 
whether there exist or not other massive deformations of type IIA. 

In this paper we shall address these questions 
by working in  ten-dimensional IIA superspace. 
The starting point of our search is the supertorsion Bianchi identities (BI) 
--which every supersymmetric system should satisfy.  
In this
purely-geometric approach 
no form superfields are introduced by hand, unlike in the usual 
superspace formulation of IIA supergravity \cite{gate}; the 
field-strengths   of the various supergravity forms `sit' inside
the components of the torsion. 
We do not make any further 
assumptions other than that any deviation from massless IIA should 
appear at canonical mass dimension one or higher. I.e. we demand that 
up to dimension  one-half, the supertorsion 
components are (equivalent to) those of massless IIA supergravity. 
As there are formulations of the latter (see for example \cite{cede}) 
in which the scalar superfield does {\it not} appear explicitly in the 
torsion components of dimension zero or one-half, 
we shall assume that the field content 
at dimension zero consists of {\it at most} one scalar superfield, 
while the field content at  dimension one-half consists of
 one chiral and one antichiral  Majorana spinor superfields.

In fact, an interesting feature
of  the formulation presented here is that
the scalar superfield at dimension zero 
(whose lowest component can be identified with the dilaton), present in
both Romans and massless IIA supergravities, 
is {\it not} introduced
from the outset but it
arises as a `potential' for the spinor superfields at dimension one-half: 
its existence follows from a certain integrability condition
implied by the BI's.
In the case of HLW supergravity
the aforementioned integrability condition fails and
such a `potential' does not exist. 
More specifically: as we show at the end of section \ref{dimension2}, both 
in the case of massless IIA and in Romans supergravity, one can construct 
a {\it closed} one-form superfield whose lowest component 
is identified with the spinor superfields at dimension one-half. 
In a topologically nontrivial spacetime this one-form may not be exact, 
a fact which leads to a topological modification of massless IIA. 
This is the ten-dimensional version of the possibility to modify 
ordinary eleven-dimensional supergravity to MM-theory.

We generally expect that
taking the supertorsion Bianchi identities as the starting point,
should allow for more freedom than starting with superforms. Evidence for this
was recently provided in \cite{cgnt}, where the supertorsion BI's 
of eleven-dimensional supergravity where solved at first 
order in a deformation
parameter related to the Planck length\footnote{The first supersymmetric 
deformation
occurs at order $l^3$ and it is of topological nature \cite{t}. 
The next deformation,
related to the $R^4$ superinvariant, is expected to occur at order $l^6$.}.
Although it has been shown that the 
four-form formulation of 11d supergravity implies the supertorsion formulation 
\cite{ht}\footnote{This was shown in \cite{ht} to first order in
the deformation parameter. In the undeformed theory, this
fact had been previously pointed out in \cite{cgnn, nr}.}, 
it was realized in \cite{cgnt} that the converse may not be true.

The main result of this paper can be stated as follows.   
Depending on the values of two scalar superfields  
($L$, $L'$ of equation (\ref{llprime}) below) arising at 
canonical mass dimension one, there exist exactly 
two massive deformations of IIA supergravity:  
Romans supergravity and HLW supergravity.

In the following section we introduce IIA superspace and establish our notation and conventions. 
We also examine the possible 
field redefinitions, in preparation for the analysis of the BI's in section \ref{bi}.
The reader who is not interested in the derivation of the final result, may skip directly 
to section \ref{summary} where the outcome of the analysis of the BI's is summarized. 
We conclude in section \ref{conclusions} with some possible future directions. 
The appendix contains our conventions on certain gamma-traceless projections used in section 
\ref{bi}.

\section{General setup}

\subsection{Type IIA superspace}

Let us begin by introducing the usual superspace machinery of vielbein ($E^A$), 
connection ($\Omega_{A}{}^B$), torsion ($T^A$) and curvature ($R_A{}^{B}$), via
\beal
T^A&=DE^A\nn\\
R_A{}^B&=d\Omega_A{}^B+\Omega_A{}^C\wedge \Omega_C{}^B~.
\end{align}
A flat superindex is denoted by a capital Latin letter from the beginning of the alphabet 
and stands for both bosonic ($a$) and fermionic ($\una$) indices. Underlined Greek  
indices from the beginning of the alphabet stand for flat fermionic indices of both chiralities. 
For example:
\beal
S^{\una}:=(S^{\al}, S_{\al})~,
\end{align}
where $S^{\al}$ is chiral and $S_{\al}$ is antichiral. Note that we never raise/lower 
chiral fermionic indices, so that the position of the index denotes a definite chirality. 
In IIA superspace the spinor part of the vielbein 
contains both chiralities: $E^{\una}:=(E^{\al}, E_{\al})$. 

The torsion and curvature satisfy the Bianchi identities 
\beal
DT^A&=E^BR_B{}^A\nn\\
DR_A{}^B&=0~.
\label{r}
\end{align}
If the connection is Lorentzian, as we assume to be the case in the present 
paper, the second BI follows from the first \cite{drag}.  Hence we need only 
analyze the first of equations (\ref{r}), i.e. the supertorsion BI.

\subsection{Field redefinitions}
\label{fieldredefinitions}

Before coming to the analysis of the BI's, let us examine 
what the possible field redefinitions are. 
We have at our disposal vielbein and connection redefinitions which can be used to 
gauge-fix some of the torsion components. Specifically, let $h_{A}{}^B:=E_A{}^M\delta E_M{}^B$ 
and $\Delta_{AB}{}^C:=E_A{}^M\delta\Omega_{MB}{}^C$. 
Under $E_M{}^A\rightarrow E_M{}^A+\delta E_M{}^A$, 
$\Omega_{MB}{}^C\rightarrow \Omega_{MB}{}^C+\delta\Omega_{MB}{}^C$ the torsion transforms as 
$T_{AB}{}^C\rightarrow T_{AB}{}^C+\delta T_{AB}{}^C$ where
\beal
\delta T_{AB}{}^C=
2\Delta_{[AB\}}{}^C+2D_{[A}h_{B\}}{}^C-2h_{[A|}{}^DT_{D|B\}}{}^C +T_{AB}{}^D h_D{}^C
~.
\label{redefs}
\end{align}
Let us analyze the possible field redefinitions in the order of 
increasing canonical mass dimension.

$\bullet$ Dimension $0$

We search for supergravities which, in the  massless limit, reduce to massless type IIA. 
Therefore, as explained in the introduction, 
we shall assume that the field content at dimension zero consists of {\it at most} 
a scalar superfield ($\phi$). Hence, the most general 
form of the dimension-zero torsion components is
\beal
T_{\al\bn}{}^c&=c_1(\ga^c)_{\al\bn} \nn\\
T^{\al\bn c}&=c_2(\ga^c)^{\al\bn}\nn\\
T_{\al}{}^{\bn c}&=0~.
\end{align}
If a dimension-zero scalar superfield $\phi$ exists, 
$c_1$, $c_2$ can be arbitrary functions of $\phi$. From (\ref{redefs}) we see that we  
can use two independent 
linear combinations of $h_{\al}{}^{\bn}$,  $h^{\al}{}_{\bn}$, $h_{a}{}^b$ to 
set $T_{\al\bn}{}^c=-i(\ga^c)_{\al\bn} $, 
$T^{\al\bn c}=-i(\ga^c)^{\al\bn}$. Note that in the case where there 
exists no scalar superfield $\phi$ at dimension zero, the remaining 
linear $h$-combination is a constant. 
Therefore it {\it cannot} be used as a redefinition at dimension one-half or higher,  
as $Dh$ vanishes.

$\bullet$ Dimension $\frac{1}{2}$

At canonical mass dimension one-half the field content consists of one right-handed and one  
left-handed Majorana spinor superfield $\mu^{\al}$, $\la_{\al}$, respectively
\footnote{Here as well one can imagine the possibility 
that the dimension one-half spinors may be `eaten' by a 
higher-dimensional superfield. However it is not difficult to see that 
in this case the Bianchi identities set 
all the torsion components to zero.}. In the massless 
IIA limit we have $\la_{\al}=D_{\al}\phi$, $\mu^{\al}=D^{\al}\phi$. As we have already 
remarked, we shall not assume this 
to be the case {\it a priori}.  
The most general form of the 
vielbein and connection redefinitions is
\beal
\Delta_{\al b}{}^c&=d_1 (\ga_b{}^c\la)_{\al}\nn\\
\Delta^{\al}{}_b{}^c&=d_2 (\ga_b{}^c\mu)^{\al}
\label{redsa}
\end{align}
and
\beal
h_a{}^{\al}&=f_1(\ga_a\la)^{\al}\nn\\
h_{a\al}&=f_2(\ga_a\mu)_{\al}
\label{redsb}
\end{align}
respectively. Note that $\Delta_{\una\bn}{}^{\ga}$ is not 
independent, but is related to  $\Delta_{\una b}{}^c$ via  
the Lorentz condition. The most general form of the torsion component $T_{\al b}{}^c$  reads
\beal
T_{\al b}{}^c&=e_1\delta_b{}^c\la_{\al}+e_2 (\ga_b{}^c\la)_{\al}
\end{align}
and similarly for $T^{\al}{}_{b}{}^c$. Hence we can use the field redefinitions 
(\ref{redsa}, \ref{redsb}) to set $T_{\al b}{}^c=T^{\al}{}_{b}{}^c=0$. 

In the case where there exists no scalar superfield at zero dimension, 
this is all we can do in the way of gauge-fixing. 
However, when $\phi$ exists $T_{\al\bn}{}^\ga$ 
can also be partially gauge-fixed as follows. The most general 
form of $T_{\al\bn}{}^\ga$  is
\beal
T_{\al\bn}{}^\ga&=g_1\la_{(\al}\de_{\bn)}{}^{\ga}+g_2(\ga^e)_{\al\bn}(\ga_e\la)^{\ga}~.
\label{kook}
\end{align}
As we can see from 
(\ref{redefs}) and the fact that $D_{\al}\phi=\la_{\al}$, the remaining 
independent linear 
combination of $h_{\al}{}^{\bn}$,  $h^{\al}{}_{\bn}$, $h_{a}{}^b$ can be used 
to gauge-fix the coefficient $g_1(\phi)$ in (\ref{kook}). 

 $\bullet$ Dimension $1$

As usual, $\Delta_{ab}{}^c$ can be used to set $T_{ab}{}^c=0$.

To summarize the results of this subsection: the 
field redefinitions can be used to set
\beal
&T_{\al\bn}{}^c=-i(\ga^c)_{\al\bn} \nn\\
&T^{\al\bn c}=-i(\ga^c)^{\al\bn}\nn\\
&T_{\una b}{}^c, ~T_{ab}{}^c=0~.
\end{align}
If, in addition, there is a scalar superfield $\phi$ such that 
$D_{\al}\phi=\la_{\al}$, then 
$T_{\al\bn}{}^{\ga}$ can also be partially gauge-fixed, as explained below (\ref{kook}).

\section{Analysis of the torsion Bianchi identities}
\label{bi}

We are now ready to come to the solution of 
the torsion Bianchi identities. We shall proceed 
systematically, in increasing order 
of canonical mass dimension. 
 The readers who are not interested in the 
details of this analysis, may skip directly to section \ref{summary} 
where the final result is summarized.

\subsection{Dimension-$\frac{1}{2}$ BI}

Taking the discussion of section \ref{fieldredefinitions} into account 
and imposing the Lorentz condition, 
the dimension one-half torsion Bianchi identity reads
\beal
T_{(\una\ub|}{}^{\une}T_{\une|\uc)}{}^e=0~.
\label{b121}
\end{align}
We distinguish the following cases.

$\bullet$  Case 1: $(\una,\ub,\uc)=(\al, \bn, \ga)$

The Bianchi identity (\ref{b121}) reads
\beal
T_{(\al\bn|}{}^{\e}(\ga^e)_{\e|\ga)}=0~.
\end{align}
Substituting (\ref{kook}) in the equation above 
and using the identity 
\beal
\ga_{(\al\bn}^e\delta_{\ga)}^{\delta}=(\ga_f)_{(\al\bn}(\ga^{ef})_{\ga)}{}^{\delta}~,
\end{align}
we obtain
\beal
T_{\al\bn}{}^\ga&=g_1\Big\{
\la_{(\al}\de_{\bn)}{}^{\ga}-\frac{1}{2}(\ga^e)_{\al\bn}(\ga_e\la)^{\ga}
\Big\}~.
\end{align}

$\bullet$  Case 2: $(\una,\ub,\uc)=(\upa,\upb,\upc)$

Similarly to the previous case we obtain
\beal
T^{\al\bn}{}_\ga&=g_2\Big\{
\mu^{(\al}
\de^{\bn)}{}_{\ga}-\frac{1}{2}(\ga_e)^{\al\bn}(\ga^e\mu)_{\ga}\Big\}~.
\end{align}

$\bullet$  Case 3: $(\una,\ub,\uc)=(\al, \bn, \upc)$

The Bianchi identity (\ref{b121}) reads
\beal
2T_{(\al|}{}^{\ga\e}(\ga^e)_{\e|\bn)}
+T_{\al\bn\e}(\ga^e)^{\e\ga}=0~.
\end{align}
We expand the torsion components above as follows
\beal
T_{\al\bn\ga}&=g_3(\ga^e)_{\al\bn}(\ga_e\mu)_\ga\nn\\
T_\al{}^{\bn\ga}&=
g_4 \de_\al{}^\ga\mu^\bn +g_5(\ga^{e_1e_2})_\al{}^\ga(\ga_{e_1e_2}\mu)^\bn
+g_6(\ga^{e_1\dots e_4})_\al{}^\ga(\ga_{e_1\dots e_4}\mu)^\bn\nn\\
~.
\end{align}
In terms of irreducible representations the Bianchi identity decomposes as\footnote{
We are using the Dynkin notation for the complex cover $D_5$ of $SO(1,9)$. I.e. 
(00000), (10000), (01000), (00100), (00011) denote a scalar, vector, two-form, three-form, 
four-form respectively. Self-dual, anti-self-dual five-forms are denoted by 
(00002), (00020) respectively and chiral, anti-chiral spinors are denoted by 
(00010), (00001). Similarly, an irreducible (gamma-traceless) chiral vector-spinor 
is denoted by (10010), a chiral two-form spinor by (01010), etc.}
$$
(00001)^{2\otimes_s}\otimes(00010)\otimes(10000)\sim3(00010)\oplus\dots
$$
where the ellipses stand for irreducible representations which drop out 
of the BI. Hence the BI imposes at most three independent conditions on the 
coefficients $g_3,\dots g_6$. These conditions can be obtained by contracting with the
 three independent structures 
$(\ga^e)^{\al\bn}\delta_{\ga}^{\e}$, $(\ga_f)^{\al\bn}(\ga^{ef})_{\ga}{}^{\e}$ and  
$(\ga^{ef_1\dots f_4})^{\al\bn}(\ga_{f_1\dots f_4})_{\ga}{}^{\e}$. 
We thus obtain $g_4=-\frac{1}{2}g_3$, $g_5=\frac{1}{4}g_3$ and 
$g_6=0$, so that
\beal
T_{\al\bn\ga}&=g_3(\ga^e)_{\al\bn}(\ga_e\mu)_\ga\nn\\
T_\al{}^{\bn\ga}&=-\frac{1}{2}g_3\Big\{
\de_\al{}^\ga\mu^\bn -\frac{1}{2}(\ga^{ef})_\al{}^\ga(\ga_{ef}\mu)^\bn
\Big\}
~.
\end{align}

$\bullet$  Case 4: $(\una,\ub,\uc)=(\al, \upb, \upc)$

Similarly to the previous case, we obtain
\beal
T^{\al\bn\ga}&=g_4(\ga_e)^{\al\bn}(\ga^e\la)^\ga\nn\\
T^\al{}_{\bn\ga}&=-\frac{1}{2}g_4 \Big\{
\de^\al{}_\ga\la_\bn -\frac{1}{2}(\ga^{ef})^\al{}_\ga(\ga_{ef}\la)_\bn
\Big\}
~.
\end{align}

\subsection{Dimension-1 BI}
\label{dimension1}

For the gamma-matrix manipulations of this and the remaining subsections, we have found 
\cite{gran} extremely useful. We have also made use of \cite{lie} in evaluating 
tensor products of representations.

Let us expand the spinor derivatives of the dimension-$\frac{1}{2}$ spinor 
superfields as follows
\beal
D_{\al}\la_\bn&=K_e{}(\ga^e)_{\al\bn}+K_{efg}{}(\ga^{efg})_{\al\bn}\nn\\
D_\al\mu^\bn&=L{}\de_\al{}^\bn+L_{ef}{}(\ga^{ef})_\al{}^\bn
+L_{efgh}{}(\ga^{efgh})_\al{}^\bn
\label{uiy}
\end{align}
and
\beal
D^{\al}\la_\bn&=
{L'}{}\de^\al{}_\bn+{L'}^{ef}{}(\ga_{ef})^\al{}_\bn
+{L'}^{efgh}{}(\ga_{efgh})^\al{}_\bn\nn\\
D^\al\mu^\bn&={K}^{\prime e}{}(\ga_e)^{\al\bn}
+{K}^{\prime efg}{}(\ga_{efg})^{\al\bn}~.
\end{align}
We also expand the dimension-one torsion as follows:
\beal
T_{ab}{}^c&=0\nn\\
T_{a\al}{}^\bn&= \de_\al{}^\bn V^1_a
+(\ga_a{}^e)_\al{}^\bn V^2_e   
+(\ga_a{}^{fgh})_\al{}^\bn H^1_{fgh}+(\ga^{fg})_\al{}^\bn H^2_{afg}
\nn\\
T_a{}^{\al}{}_{\bn}&= \de^\al{}_\bn V^{\prime1}_a
+(\ga_a{}^e)^\al{}_\bn V^{\prime2}_e 
+(\ga_a{}^{fgh})^\al{}_\bn H^{\prime 1}_{fgh}
+(\ga^{fg})^\al{}_\bn H^{\prime 2}_{afg}
\nn\\
T_{a\al\bn}&=
S (\ga_a)_{\al\bn} +(\ga_a{}^{ef})_{\al\bn}F^1_{ef} 
+(\ga^{e})_{\al\bn}F^2_{ae}
+(\ga_a{}^{efgh})_{\al\bn}G^1_{efgh} 
+(\ga^{efg})_{\al\bn}G^2_{aefg}
\nn\\
T_a{}^{\al\bn}&=
S' (\ga_a)^{\al\bn} +(\ga_a{}^{ef})^{\al\bn}{F}^{\prime1}_{ef} 
+(\ga^{e})^{\al\bn}{F}^{\prime2}_{ae}
+(\ga_a{}^{efgh})^{\al\bn}{G}^{\prime1}_{efgh} 
+(\ga^{efg})^{\al\bn}{G}^{\prime2}_{aefg}
~.
\label{t1}
\end{align}
The superfields appearing on the right-hand-side of (\ref{t1})
are all forms. We can see that there can be no hooks in the 
above expansions, for the following reason. Assuming there is 
a hook superfield ($U$) at dimension one, we can expand 
$U=m U_{(0)}+ U_{(1)}$, where $m$ is a mass parameter, so that  
$U_{(0)}$ is of canonical mass dimension zero, 
$ U_{(1)}$ is of canonical mass dimension one and 
does not depend on $m$. Taking the massless $m\rightarrow 0$ 
limit we see that 
$ U_{(1)}$ has to vanish, as no hook superfields can appear 
in the torsion components of massless IIA. Also $U_{(0)}$ has to vanish, since 
at dimension zero there can exist at most a scalar superfield. 
Using the same argument we can see that there can be no 
five-forms in the expansions (\ref{uiy}-\ref{t1}).

Taking the Lorentz condition 
into account, the Bianchi identities at dimension one read
\beal
R_{\una\ub} {}_{cd}
       &= 2T_{c(\una|}{}^{\une} T_{\une|\ub)d} ~.
\label{bi1.1}
\end{align}
%
%
%
and
\beal
R_{(\una\ub}{}_{\uc)}{}^{\und}&= D_{(\una}T_{\ub\uc)}{}^{\und}
         +T_{(\una\ub|}{}^eT_{e|\uc)}{}^{\und}+T_{(\una\ub|}{}^{\une}
         T_{\une|\uc)}{}^{\und} ~.
\label{bi1.2}
\end{align}
Let us analyze (\ref{bi1.1}) first. We distinguish the following cases.

$\bullet$  Case 1: $(\una,\ub)=(\al,\bn)$

Demanding that the right-hand-side of (\ref{bi1.1}) be antisymmetric 
in $c,d$ implies
\be
V^1_a=V^2_a=0
\end{equation}
and the curvature is given by
\beal
R_{\al\bn} {}_{cd}=2i(\ga_{cd}{}^{efg})_{\al\bn}H^1_{efg}
+4i(\ga^{e})_{\al\bn}H^2_{cde}~.
\label{r1}
\end{align}

$\bullet$  Case 2: $(\una,\ub)=(\upa,\upb)$

Similarly to the previous case we get
\be
V^{\prime1}_a=V^{\prime2}_a=0
\end{equation}
and
\beal
R^{\al\bn} {}_{cd}=2i(\ga_{cd}{}^{efg})^{\al\bn}H^{\prime 1}_{efg}
+4i(\ga^{e})^{\al\bn}H^{\prime 2}_{cde}~.
\label{r3}
\end{align}

$\bullet$  Case 3: $(\una,\ub)=(\al,\upb)$

Demanding that the right-hand-side of (\ref{bi1.1}) be antisymmetric 
in $c,d$ in this case implies
\beal
S&=-S'\nn\\
F^1&={F}^{\prime1}\nn\\
F^2&={F}^{\prime2}\nn\\
G^1&=-{G}^{\prime1}\nn\\
G^2&=-{G}^{\prime2}
\label{lamb}
\end{align}
and the curvature is given by
\beal
R_{\al}{}^{\bn} {}_{cd}=-2i\Big\{
(\ga_{cd})_{\al}{}^\bn S+(\ga_{cd}{}^{ef})_{\al}{}^\bn F^1_{ef}
+\de_{\al}{}^\bn F^2_{cd} +
(\ga_{cd}{}^{efgh})_{\al}{}^\bn G^1_{efgh}
+3(\ga^{fg})_{\al}{}^\bn G^2_{cdfg}
\Big\}~.
\label{r2}
\end{align}

Let us now come to (\ref{bi1.2}). We distinguish the following cases.

$\bullet$  Case 1.1: $(\una,\ub, \uc, \und)=(\al,\bn, \ga, \de)$

We shall work out this case in 
some detail in order to illustrate the general procedure. 
Taking the Lorentz condition into account, we obtain
\beal
&(\ga^{e})_{(\al\bn}(\ga_e{}^{fgh})_{\ga)}{}^\de 
\Big\{ iH^1_{fgh}-\frac{1}{2}g_1K_{fgh}
-\frac{1}{96}(g_1^2+\frac{\dot{g}_1}{2}) (\la\ga_{fgh}\la)\Big\}\nn\\
&+(\ga^{e})_{(\al\bn}(\ga^{fg})_{\ga)}{}^\de \Big\{2iH^2_{efg}
+\frac{3}{2}g_1 K_{efg}
-\frac{1}{4} (\mu\ga_{efg}\mu)
-\frac{3}{96}(g_1^2-\frac{\dot{g}_1}{2})(\la\ga_{efg}\la)
\Big\}\nn\\
&+\frac{i}{2}(\ga_{cd}{}^{efg})_{(\al\bn}(\ga^{cd})_{\ga)}{}^\de H^1_{efg}
-\frac{1}{2}g_1 K^f\Big\{ 
(\ga_f)_{(\al\bn}\delta_{\ga)}^{\delta}
+(\ga^e)_{(\al\bn}(\ga_{ef})_{\ga)}{}^{\delta}  \Big\}=0~,
\label{klug}
\end{align}
where we have allowed for the possibility that $\la_{\al}={D}_{\al}\phi$ and we have 
set $\dot{g}_1:=\frac{d}{d\phi}g_1$. 
Note that the vector drops out of the BI due to the identity 
$$
(\ga_f)_{(\al\bn}\delta_{\ga)}^{\delta}=-(\ga^e)_{(\al\bn}(\ga_{ef})_{\ga)}{}^{\delta}~.
$$
Moreover, one can see that there is a unique 
three-form in the decomposition of the 
tensor product of a chiral spinor and the symmetrized tensor product 
of three antichiral spinors:
\beal
(00010)\otimes (00001)^{3\otimes_s}\sim 1(10000)\oplus 1(00100)\oplus\dots
\end{align}
Therefore, equation (\ref{klug}) imposes at most one linear equation 
on the three-forms. Contracting both sides 
of (\ref{klug}) with $(\ga_{[a})^{\al\bn}(\ga_{bc]})_\de{}^\ga$  
in order to saturate the spinor indices, we obtain\footnote{
One can verify that the same equation is obtained 
by contracting (\ref{klug}) with, for example, $(\ga^{e})^{\al\bn}(\ga_{eabc})_\de{}^\ga$. 
Here and in the rest of the analysis of the BI's, we have applied many more contractions 
than the number of independent ones. This is not strictly-speaking necessary, but the 
`redundant ' contractions serve as useful consistency checks.}
\beal
48i H^1_{abc}-8i H^2_{abc}-12g_1K_{abc}+(\mu\ga_{abc}\mu)
+\frac{1}{16}(g_1^2-2\dot{g}_1)(\la\ga_{abc}\la)=0~.
\label{1}
\end{align}

$\bullet$  Case 1.2: $(\una,\ub, \uc, \und)=(\al,\bn, \ga, \upd)$

In terms of irreducible representations the BI decomposes as
\beal
(00001)\otimes (00001)^{3\otimes_s}\sim 1(01000)\oplus 1(00011)\oplus\dots
\end{align}
Hence the BI imposes at most one linear equation on the two-forms and one on the four-forms. 
Contracting with $(\ga_{[a})^{\al\bn}(\ga_{b]})^{\de\ga}$ 
we obtain 
\beal
0=L_{ab}+\frac{1}{4}(\mu\ga_{ab}\la)
-\frac{i}{2}F^1_{ab}+\frac{i}{4}F^2_{ab}
~.
\label{tri}
\end{align}
Similarly, contracting with $(\ga_{[a})^{\al\bn}(\ga_{bcd]})^{\de\ga}$ 
we obtain 
\beal
0=L_{abcd}-\frac{i}{2}G^1_{abcd}+\frac{i}{8}G^2_{abcd}
~.
\label{trii}
\end{align}

$\bullet$  Case 2.1: $(\una,\ub, \uc, \und)=(\al,\bn, \upc, \de)$

In terms of representations, the BI decomposes as
\beal
(00010)^{2\otimes}\otimes(00001)^{2\otimes_s}
\sim 2(00000)\oplus 4(01000)\oplus 5(00011)\oplus \dots
\label{repr}
\end{align}
I.e. the BI imposes at most two constraints on the scalars, four on the two-forms and 
five on the four-forms. Let us examine each representation in turn. 

{\it Scalars}:

Using the independent structures 
$(S_1)^{\al\bn}_{\ga\de}:=(\ga^a)^{\al\bn}(\ga_a)_{\ga\de}$ 
and $(S_2)^{\al\bn}_{\ga\de}:=(\ga^{a_1\dots a_5})^{\al\bn}(\ga_{a_1\dots a_5})_{\ga\de}$ 
to saturate the spinor indices\footnote{ 
The fact that these structures are independent is seen as follows. 
Contracting the equation 
$$A (\ga^a)_{\al\bn}(\ga_a)^{\ga\de}+ B(\ga^{a_1\dots a_5})_{\al\bn}(\ga_{a_1\dots a_5})^{\ga\de}=0$$ 
with $S_1$, $S_2$, we obtain $A=B=0$. Note that there are exactly two independent structures 
as follows from the representation-theoretic analysis (\ref{repr}).}, 
we obtain
\beal
S=-\frac{2i}{5}L-\frac{ig_1}{10}L'+i\Big(\frac{1}{20}-\frac{11g_1}{80}
-\frac{\dot{g}_1}{160}\Big)(\mu\la)~.
\label{gu}
\end{align}
Hence in this case both independent contractions yield the same equation.

{\it Two-forms}: 

Contracting with the four independent structures 
$(\ga_{[a})^{\al\bn}(\ga_{b]})_{\ga\de}$, 
$(\ga^f)^{\al\bn}(\ga_{fab})_{\ga\de}$, 
$(\ga^{abfgh})^{\al\bn}(\ga_{fgh})_{\ga\de}$ 
and $(\ga^{[a|efgh})^{\al\bn}(\ga^{|b]}{}_{efgh})_{\ga\de}$, we get 
\beal
0&=\frac{g_1}{6}L'_{ab}+iF_{ab}^1+\Big(\frac{1}{8}+\frac{g_1}{32}
+\frac{\dot{g}_1}{192}\Big)(\mu\ga_{ab}\la)\nn\\
0&=4L_{ab}+\frac{g_1}{3}L'_{ab}+iF_{ab}^2+\Big(\frac{5}{4}+\frac{g_1}{16} 
+\frac{\dot{g}_1}{96}\Big)(\mu\ga_{ab}\la)~.
\label{try}
\end{align}
Remarkably, all four independent contractions yield only the two constraints above. 
We also remark that equations (\ref{try}) imply (\ref{tri}), which is therefore not independent.

{\it Four-forms}: 

Contracting with the five 
independent structures 
$(\ga^{[a})^{\al\bn}(\ga^{bcd]})_{\ga\de}$, 
$(\ga^{e})^{\al\bn}(\ga_e{}^{abcd})_{\ga\de}$, 
$(\ga^{abcde})^{\al\bn}(\ga_e)_{\ga\de}$, 
$(\ga^{[abc|ef})^{\al\bn}(\ga^{|d]}{}_{ef})_{\ga\de}$ and 
$(\ga^{[ab|efg})^{\al\bn}(\ga^{cd]}{}_{efg})_{\ga\de}$ 
we get
\beal
0&=2L_{abcd}-\frac{g_1}{2}L'_{abcd}+iG^1_{abcd}
+\Big(\frac{1}{96}+\frac{g_1}{384}-\frac{\dot{g}_1}{768}\Big)(\mu\ga_{abcd}\la)\nn\\
0&=16L_{abcd}-2g_1 L'_{abcd}+iG^2_{abcd}
+\Big(\frac{1}{24}+\frac{g_1}{96}-\frac{\dot{g}_1}{192}\Big)(\mu\ga_{abcd}\la)~.
\label{mn}
\end{align}
I.e. all five independent contractions yield only the two constraints above. 
Moreover equations (\ref{mn}) imply (\ref{trii}).

$\bullet$  Case 2.2: $(\una,\ub, \uc, \und)=(\al,\bn, \upc, \upd)$

The BI decomposes as
\beal
(00001)^{2\otimes_s}\otimes(00001)\otimes(00010)
\sim 3(10000)\oplus 4(00100)\oplus \dots
\label{repri}
\end{align}
Therefore the BI will yield at most three constraints on the vectors and four on the three-forms. 
It can be seen that the vector part of the BI is equivalent to
\beal
K_a=K_a'~.
\label{po1}
\end{align}
Moreover, contracting with the four independent structures 
$(\ga^{ef})_\ga{}^\de(\ga_{ef}{}^{abc})^{\al\bn}$, 
$(\ga^{[a| efg})_\ga{}^\de(\ga_{efg}{}^{|bc]})^{\al\bn}$, 
$(\ga^{[ab})_\ga{}^\de(\ga^{c]})^{\al\bn}$ and 
$(\ga^{abce})_\ga{}^\de(\ga_{e})^{\al\bn}$ we obtain
\beal
0&=
iH^1_{abc}+K_{abc}+\Big(\frac{1}{48}+\frac{g_1}{192}  \Big)(\la\ga_{abc}\la)
\nn\\
0&=
iH^2_{abc}+iH^{\prime 2}_{abc}
-3K_{abc}-3K'_{abc}
+\Big(\frac{3}{16}+\frac{7g_1}{64}\Big)(\mu\ga_{abc}\mu)
+\Big(\frac{3}{16}+\frac{7g_2}{64}\Big)(\la\ga_{abc}\la)\nn\\
0&=
iH^{\prime 1}_{abc}+K'_{abc}+
\Big(\frac{1}{48}+\frac{g_2}{192}\Big)(\mu\ga_{abc}\mu)~.
\label{po}
\end{align}

\vfill\break

$\bullet$  Case 3.1: $(\una,\ub, \uc, \und)=(\al,\upb, \upc, \de)$

In terms of representations, this is the same as case 2.2 above. The equations 
which follow from this BI turn out to be identical to (\ref{po1}, \ref{po}).

$\bullet$  Case 3.2: $(\una,\ub, \uc, \und)=(\al,\upb, \upc, \upd)$

In terms of representations, this is the same as case 2.1. 
Proceeding similarly, we obtain the following equations

{\it Scalars}: 

\beal
S=\frac{2i}{5}L'+\frac{ig_2}{10}L+i\Big(\frac{1}{20}-\frac{11g_2}{80}
-\frac{\dot{g}_2}{160}\Big)(\mu\la)~.
\label{gub}
\end{align}

{\it Two-forms}:

\beal
0&=\frac{g_2}{6}L_{ab}+iF_{ab}^1+\Big(\frac{1}{8}+\frac{g_2}{32}
+\frac{\dot{g}_2}{192}\Big)(\mu\ga_{ab}\la)\nn\\
0&=4L'_{ab}+\frac{g_2}{3}L_{ab}+iF_{ab}^2+\Big(\frac{5}{4}+\frac{g_2}{16} 
+\frac{\dot{g}_2}{96}\Big)(\mu\ga_{ab}\la)~.
\label{tryb}
\end{align}

{\it Four-forms}:

\beal
0&=2L'_{abcd}-\frac{g_2}{2}L_{abcd}-iG^1_{abcd}
-\Big(\frac{1}{96}+\frac{g_2}{384}+\frac{\dot{g}_1}{768}\Big)(\mu\ga_{abcd}\la)\nn\\
0&=16L'_{abcd}-2g_2 L_{abcd}-iG^2_{abcd}
-\Big(\frac{1}{24}+\frac{g_2}{96}+\frac{\dot{g}_2}{192}\Big)(\mu\ga_{abcd}\la)~.
\end{align}

$\bullet$  Case 4.1: $(\una,\ub, \uc, \und)=(\upa,\upb, \upc, \de)$

In terms of representations, this is the same as case 1.2. Proceeding similarly 
we obtain the following equations
\beal
0=L'_{ab}+\frac{1}{4}(\mu\ga_{ab}\la)
-\frac{i}{2}F^1_{ab}+\frac{i}{4}F^2_{ab}
\label{trib}
\end{align}
and
\beal
0=L'_{abcd}+\frac{i}{2}G^1_{abcd}-\frac{i}{8}G^2_{abcd}
~.
\label{triib}
\end{align}

\vfill\break

$\bullet$  Case 4.2: $(\una,\ub, \uc, \und)=(\upa,\upb, \upc, \upd)$

In terms of representations, this is the same as case 1.1. 
Proceeding similarly we obtain
\beal
48i H^{\prime 1}_{abc}-8i H^{\prime 2}_{abc}
-12g_2K'_{abc}+(\la\ga_{abc}\la)
+\frac{1}{16}(g_2^2-2\dot{g}_2)(\mu\ga_{abc}\mu)=0~.
\label{3}
\end{align}

The system of equations above imply in particular
\beal
L_{ab}&=L'_{ab}\nn\\
L_{abcd}&=-L'_{abcd}
\label{inta}
\end{align}
and
\beal
g_1-g_2&=0\nn\\
(g_1+4)(L+L')&=0~. 
\label{intb}
\end{align}
Conditions (\ref{po1}, \ref{inta}) 
if supplemented with $L=-L'$, would imply ${D}_{\al}\mu^{\bn}=-{D}^{\bn}\la_{\al}$, 
$(\ga_a)^{\al\bn}{D}_{\al}\la_{\bn}
=(\ga_a)_{\al\bn}{D}^{\al}\mu^{\bn}$. 
The latter two equations 
are solved for $\la_{\al}={D}_{\al}\phi$, $\mu^{\al}={D}^{\al}\phi$. 
We shall see in the following that in the case $g_1=-4$, $L=-L'$, the existence of $\phi$ 
is indeed implied by the  higher-dimension BI's. But we have seen 
in section \ref{fieldredefinitions} that in this case 
the coefficient $g_1$ can be shifted by field redefinitions. Hence the case $g_1=-4$, $L=-L'$ is 
in fact equivalent to the case $g_1\neq -4$, $L=-L'$. 
In the following we shall set $g_1=-4$, so that we can treat both $L=-L'$ and $L\neq -L'$ 
cases simultaneously. Note that if $L\neq -L'$, there cannot exist a $\phi$ such that 
$\la_{\al}={D}_{\al}\phi$, $\mu^{\al}={D}^{\al}\phi$.

\subsection{Dimension-$\frac{3}{2}$ BI}

Taking into account our gauge-fixing in section \ref{fieldredefinitions}, 
the Bianchi identities at canonical mass dimension three-half read
\beal
2R_{\una[bc]}{}^d=T_{bc}{}^{\une}T_{\une\una}{}^d
\label{bi321}
\end{align}
and
\beal
2R_{e(\una\ub)}{}^{\und}=2D_{(\una}T_{\ub)e}{}^{\und}+D_eT_{\una\ub}{}^{\und}
+T_{\una\ub}{}^fT_{fe}{}^{\und}+T_{\una\ub}{}^{\une}T_{\une e}{}^{\und}
+2T_{e(\una}{}^{\une}T_{\une|\ub)}{}^{\und}~.
\label{bi322}
\end{align}
Equation (\ref{bi321}) can be solved for the dimension three-half supercurvature to give
\beal
R_{\una bcd}=\frac{i}{2}\Big\{  
(\ga_bT_{cd})_{\una}+(\ga_cT_{bd})_{\una} -(\ga_dT_{bc})_{\una}  \Big\}~.
\end{align}
In the following we shall expand the dimension-$\frac{3}{2}$ 
torsion into irreducible (gamma-traceless) parts as follows
\beal
T_{ab}=\widetilde{T}_{ab}
+{}\ga_{[a}\widetilde{T}_{b]}{}{}
+{}\ga_{ab}\widetilde{T}
~,
\end{align}
where we have suppressed all spinor indices. Similarly, we expand the spinor derivatives 
of the various dimension-one fields in irreducible representations, 
as follows (again suppressing spinor indices)
\beal
DL&=\widetilde{L}\nn\\
DL'&=\widetilde{L}'\nn\\
DK_a&=\widetilde{K}^{(1)}_a+\ga_a\widetilde{K}^{(1)}\nn\\
DL_{ab}&=\widetilde{L}^{(2)}_{ab}+\ga_{[a}\widetilde{L}^{(2)}_{b]}+\ga_{ab}\widetilde{L}^{(2)}\nn\\
DK_{abc}&=\widetilde{K}^{(3)}_{abc}+\ga_{[a}\widetilde{K}^{(3)}_{bc]}
+\ga_{[ab}\widetilde{K}^{(3)}_{c]}
+\ga_{abc}\widetilde{K}^{(3)}_{}\nn\\
DL_{abcd}&=\widetilde{L}^{(4)}_{abcd}+\ga_{[a}\widetilde{L}^{(4)}_{bcd]}
+\ga_{[ab}\widetilde{L}^{(4)}_{cd]}
+\ga_{[abc}\widetilde{L}^{(4)}_{d]}+\ga_{abcd}\widetilde{L}^{(4)}_{}~.
\end{align}

Let us now come to equation (\ref{bi322}). We distinguish the following cases. 

$\bullet$  Case 1.1: $(\una,\ub, \und)=(\al, \bn, \de)$

In terms of irreducible representations the BI decomposes as
$$
(10000)\otimes(00010)\otimes(00001)^{2\otimes_s}\sim 
3(00010)\oplus 4(10001)\oplus2(01010)\oplus2(00101)\oplus(00021)\oplus\dots
$$
Hence the BI imposes at most three constraints on the spinors, four on the vector-spinors, etc. 
Let us analyze each representation in turn:

{\it Spinors}

Contracting with the three independent structures $(\ga^e)^{\al\bn}\delta_{\delta}{}^{\ga}$, 
$(\ga_g)^{\al\bn}(\ga^{ge})_{\delta}{}^{\ga}$ and 
$(\ga^{eg_1\dots g_4})^{\al\bn}(\ga_{g_1\dots g_4})_{\delta}{}^{\ga}$, we obtain
\beal
\widetilde{T}&=-\frac{14i}{135}\slsh D \la+\frac{32}{45}L\mu+\frac{8}{45}L'\mu
+\frac{136}{405}L_{(2)}(\ga^{(2)}\mu)
+\frac{8}{45}L_{(4)}(\ga^{(4)}\mu)\nn\\
&-\frac{16}{45}K_{(1)}(\ga^{(1)}\la)
-\frac{2}{45}K_{(3)}(\ga^{(3)}\la)
-\frac{1}{60}(\mu\ga_{(3)}\mu)(\ga^{(3)}\la)\nn\\
\widetilde{K}^{(3)}_{}&=-\frac{i}{540}\slsh D \la+\frac{1}{180}L \mu+\frac{1}{45}L' \mu
 -\frac{13}{1620}L_{(2)}(\ga^{(2)}\mu)
+\frac{1}{180}L_{(4)}(\ga^{(4)}\mu)\nn\\
&-\frac{1}{90}K_{(1)}(\ga^{(1)}\la)
+\frac{1}{45}K_{(3)}(\ga^{(3)}\la)~,
\end{align}
where, to simplify the expressions, we have introduced the notation 
$\ga_{(p)}A^{(p)}:=\ga_{a_1\dots a_p}A^{a_1\dots a_p}$.

{\it Vector-spinors}

Contracting the BI with the four independent structures 
$\delta_a{}^e(\ga_g)^{\al\bn}(\ga^{g})_{\delta\ga}$,  
$\delta_a{}^e(\ga_{g_1\dots g_5})^{\al\bn}(\ga^{g_1\dots g_5})_{\delta\ga}$, \break 
$(\ga_a)^{\al\bn}(\ga^{e})_{\delta\ga}$, 
$(\ga_{a g_1\dots g_4})^{\al\bn}(\ga^{eg_1\dots g_4})_{\delta\ga}$ and 
projecting onto the gamma-traceless part, we obtain
\beal
\widetilde{T}_a&=-\frac{3i}{20}(\ga_a^{(1)} D_{(1)} \la)-\frac{1}{5}L_{(2)}(\ga_{a}^{(2)}\mu)
-\frac{2}{5}L_{(4)}(\ga_{a}^{(4)}\mu)\nn\\
&+\frac{1}{5}K_{(1)}(\ga_{a}^{(1)}\la)+\frac{3}{20}K_{(3)}(\ga_{a}^{(3)}\la)
+\frac{3}{160}(\mu\ga_{(3)}\mu)(\ga_{a}^{(3)}\la)\nn\\
\widetilde{K}^{(3)}_{a}&=
\frac{i}{160}(\ga_a^{(1)} D_{(1)} \la)+\frac{1}{40}L_{(2)}(\ga_{a}^{(2)}\mu)
+\frac{1}{20}L_{(4)}(\ga_{a}^{(4)}\mu)\nn\\
&-\frac{1}{40}K_{(1)}(\ga_{a}^{(1)}\la)+\frac{81}{1120}K_{(3)}(\ga_{a}^{(3)}\la)
-\frac{1}{1280}(\mu\ga_{(3)}\mu)(\ga_{a}^{(3)}\la)~.
\end{align}
The notation in the equation above is a shorthand for the projection onto the 
irreducible (gamma-traceless) vector-spinor part. Our conventions are 
explained in appendix \ref{gammatraceless}.

{\it Two-form-spinors}

Contracting with the two independent structures 
$(\ga_{[a})^{\al\bn}(\ga_{b]}{}^{e})_{\delta}{}^{\ga}$, 
$(\ga_{ab}{}^{eg_1g_2})^{\al\bn}(\ga_{g_1g_2})_{\delta}{}^{\ga}$, we obtain
\beal
\widetilde{K}^{(3)}_{ab}&=
\frac{1}{8}\widetilde{T}_{ab}+\frac{1}{108}L_{(2)}(\ga_{ab}^{(2)}\mu)
+\frac{1}{6}L_{(4)}(\ga_{ab}^{(4)}\mu)+\frac{1}{18}K_{(3)}(\ga_{ab}^{(3)}\la)
~.
\end{align}
As in the previous case, the projections onto the gamma-traceless part of the 
two-form-spinors are explained in appendix \ref{gammatraceless}.

{\it Three-form-spinors}

Contracting with the two independent structures 
$\delta_{[a}{}^e(\ga_{bc]g_1g_2g_3})^{\al\bn}(\ga^{g_1g_2g_3})_{\delta\ga}$ 
and $(\ga_{abc}{}^{eg})^{\al\bn}(\ga_{g})_{\delta\ga}$, we obtain
\beal
\widetilde{K}^{(3)}_{abc}&=-\frac{1}{84}K_{(3)}(\ga_{abc}^{(3)}\la)~.
\end{align}

{\it Four-form-spinors}

Contracting with 
$(\ga_{abcd}{}^e)^{\al\bn}\delta_{\delta}{}^{\ga}$, we obtain
\beal
\widetilde{L}^{(4)}_{abcd{}}&=0~.
\end{align}

$\bullet$ Case 1.2: $(\una,\ub, \und)=(\al, \bn, \upd)$

In terms of irreducible representations the BI decomposes as 
$$
(10000)\otimes(00001)\otimes(00001)^{2\otimes_s}\sim 
3(00001)\oplus 3(10010)\oplus3(01001)\oplus(00110)\oplus2(00012)\oplus\dots
$$
hence there will be at most three constraints on the spinors, three on the vector-spinors, etc. 
Let us examine each irreducible representation in turn:

{\it Spinors}

Contracting with the independent structures $(\ga^e)^{\al\bn}\delta_{\ga}{}^{\delta}$, 
$(\ga_g)^{\al\bn}(\ga^{ge})_{\ga}{}^{\delta}$ and 
$(\ga^{eg_1\dots g_4})^{\al\bn}(\ga_{g_1\dots g_4})_{\ga}{}^{\delta}$, we obtain
\beal
\widetilde{L}^{{}}&=\widetilde{L}^{\prime}-\frac{i}{16}\slsh D \mu
+\frac{25}{8}L \la +\frac{1}{4}L' \la+\frac{73}{24}L_{(2)}(\ga^{(2)}\la)
+\frac{9}{8}L_{(4)}(\ga^{(4)}\la)\nn\\
&-\frac{3}{8}K_{(1)}(\ga^{(1)}\mu)
-\frac{15}{16}K_{(3)}(\ga^{(3)}\mu)
-\frac{1}{64}(\la\ga_{(3)}\la)(\ga^{(3)}\mu)-\frac{135}{64}\widetilde{T}\nn\\
\widetilde{L}^{(2){}}_{}&=-\frac{3i}{320}\slsh D \mu
-\frac{21}{160}L \la +\frac{3}{80}L' \la +\frac{59}{480}L_{(2)}(\ga^{(2)}\la)
-\frac{1}{32}L_{(4)}(\ga^{(4)}\la)\nn\\
&-\frac{9}{160}K_{(1)}(\ga^{(1)}\mu)
+\frac{19}{320}K_{(3)}(\ga^{(3)}\mu)
-\frac{3}{1280}(\la\ga_{(3)}\la)(\ga^{(3)}\mu) -\frac{33}{256}\widetilde{T}   \nn\\
\widetilde{L}^{(4)}_{{}}&=-\frac{i}{3840}\slsh D \mu
-\frac{7}{1920}L \la +\frac{1}{960}L' \la -\frac{1}{3456}L_{(2)}(\ga^{(2)}\la)
-\frac{131}{40320}L_{(4)}(\ga^{(4)}\la)\nn\\
&-\frac{1}{640}K_{(1)}(\ga^{(1)}\mu)
-\frac{13}{11520}K_{(3)}(\ga^{(3)}\mu)
-\frac{1}{15360}(\la\ga_{(3)}\la)(\ga^{(3)}\mu)-\frac{1}{1024}\widetilde{T} ~.
\end{align}

{\it Vector-spinors}

Contracting with the three independent structures $\delta_a{}^e(\ga_g)^{\al\bn}(\ga^{g})^{\ga\delta}$, 
$(\ga_{ag_1\dots g_4})^{\al\bn}(\ga^{eg_1\dots g_4})^{\ga\delta}$ and \break
$(\ga_{a})^{\al\bn}(\ga^{e})^{\ga\delta}$, we obtain
\beal
\widetilde{L}^{(2){}}_{a}&=
-\frac{9i}{320}(\ga_a^{(1)} D_{(1)} \mu)-
\frac{21}{80}L_{(2)}(\ga_a^{(2)}\la)-
\frac{3}{40}L_{(4)}(\ga_a^{(4)}\la)\nn\\
&+\frac{3}{16}K_{(1)}(\ga_a^{(1)}\mu)
+\frac{63}{320}K_{(3)}(\ga_a^{(3)}\mu)
+\frac{9}{2560}(\la\ga_{(3)}\la)(\ga_a^{(3)}\mu)\nn\\
\widetilde{L}^{(4)}_{a{}}&=
-\frac{i}{1920}(\ga_a^{(1)} D_{(1)} \mu)-
\frac{1}{480}L_{(2)}(\ga_a^{(2)}\la)+
\frac{31}{1680}L_{(4)}(\ga_a^{(4)}\la)\nn\\
&-\frac{1}{480}K_{(1)}(\ga_a^{(1)}\mu)
+\frac{11}{4480}K_{(3)}(\ga_a^{(3)}\mu)
+\frac{1}{15360}(\la\ga_{(3)}\la)(\ga_a^{(3)}\mu)\nn\\
\widetilde{T}_a&=-\frac{3i}{20}(\ga_a^{(1)} D_{(1)} \mu)-\frac{1}{5}L_{(2)}(\ga_{a}^{(2)}\la)
+\frac{2}{5}L_{(4)}(\ga_{a}^{(4)}\la)\nn\\
&+\frac{1}{5}K_{(1)}(\ga_{a}^{(1)}\mu)-\frac{3}{20}K_{(3)}(\ga_{a}^{(3)}\mu)
+\frac{3}{160}(\la\ga_{(3)}\la)(\ga_{a}^{(3)}\mu)~.
\end{align}

{\it Two-form-spinors}

Contracting with the three independent structures 
$\delta_{[\al}{}^e(\ga_{b]})^{\al\bn}\delta_{\ga}{}^{\delta}$, 
$(\ga_{ab}{}^{eg_1g_2})^{\al\bn}(\ga_{g_1g_2})_{\ga}{}^{\delta}$, 
$\delta_{[\al}{}^e(\ga_{b]})^{\delta(\al}\delta_{\ga}{}^{\bn)}$
 and projecting onto the gamma-traceless part, we obtain
\beal
\widetilde{L}^{(2){}}_{ab}&=
\frac{3}{16}\widetilde{T}_{ab}
-\frac{1}{24}L_{(2)}(\ga_{ab}^{(2)}\la)-\frac{1}{4}L_{(4)}(\ga_{ab}^{(4)}\la)
-\frac{1}{8}K_{(3)}(\ga_{ab}^{(3)}\mu)\nn\\
\widetilde{L}^{(4)}_{ab{}}&=
\frac{1}{32}\widetilde{T}_{ab}
-\frac{1}{432}L_{(2)}(\ga_{ab}^{(2)}\la)
-\frac{13}{360}L_{(4)}(\ga_{ab}^{(4)}\la)
-\frac{1}{144}K_{(3)}(\ga_{ab}^{(3)}\mu)~.
\end{align}

{\it Three-form-spinors}

Contracting with $(\ga_{abc}{}^{eg})^{\al\bn}(\ga_{g})^{\ga\delta}$ we obtain
\beal
\widetilde{L}^{(4)}_{abc{}}&=
-\frac{1}{168}L_{(4)}(\ga_{abc}^{(4)}\la)
-\frac{1}{336}K_{(3)}(\ga_{abc}^{(3)}\mu)~.
\end{align}

{\it Four-form-spinors}

Contracting with the two independent structures 
$(\ga_{abcd}{}^{e})^{\al\bn}\delta_{\ga}{}^{\delta}$ and 
$(\ga_{abcd}{}^{e})^{\delta(\al}\delta_{\ga}{}^{\bn)}$, we obtain
\beal
\widetilde{L}^{(4)}_{abcd{}}&=0~.
\end{align}

$\bullet$  Case 2.1: $(\una,\ub, \und)=(\al, \upb, \de)$

In terms of irreducible representations the BI decomposes as
$$
(10000)\otimes(00001)\otimes(00010)^{2\otimes}\sim 
5(00001)\oplus 7(10010)\oplus5(01001)\oplus4(00110)\oplus2(00012)\oplus\dots
$$
Hence the BI imposes at most five constraints on the spinors, seven on the vector-spinors, etc. 
Let us analyze each representation in turn:

\vfill\break

{\it Spinors}

Contracting with the five independent structures 
$(\ga^e)_{\ga\bn}\delta_{\delta}{}^{\al}$, 
$(\ga_g)_{\ga\bn}(\ga^{eg})_{\delta}{}^{\al}$, 
$(\ga^{eg_1g_2})_{\ga\bn}(\ga_{g_1g_2})_{\delta}{}^{\al}$, \break
$(\ga_{g_1g_2g_3})_{\ga\bn}(\ga^{eg_1g_2g_3})_{\delta}{}^{\al}$ and 
$(\ga^{eg_1\dots g_4})_{\ga\bn}(\ga_{g_1\dots g_4})_{\delta}{}^{\al}$, 
we obtain
\beal
\widetilde{L}^{{}}&=\widetilde{L}^{\prime}+\frac{5i}{16}\slsh D \mu
+2L \la -\frac{1}{2}L' \la+\frac{19}{6}L_{(2)}(\ga^{(2)}\la)
+\frac{3}{2}L_{(4)}(\ga^{(4)}\la)\nn\\
&+\frac{9}{4}K_{(1)}(\ga^{(1)}\mu)
-\frac{9}{16}K_{(3)}(\ga^{(3)}\mu)
+\frac{1}{128}(\la\ga_{(3)}\la)(\ga^{(3)}\mu)\nn\\
\widetilde{L}^{(2){}}_{}&=\frac{13i}{960}\slsh D \mu
-\frac{1}{5}L \la -\frac{1}{120}L' \la +\frac{47}{360}L_{(2)}(\ga^{(2)}\la)
-\frac{1}{120}L_{(4)}(\ga^{(4)}\la)\nn\\
&+\frac{5}{48}K_{(1)}(\ga^{(1)}\mu)
+\frac{79}{960}K_{(3)}(\ga^{(3)}\mu)
-\frac{7}{7680}(\la\ga_{(3)}\la)(\ga^{(3)}\mu)\nn\\
\widetilde{L}^{(4)}_{{}}&=-\frac{i}{11520}\slsh D \mu
-\frac{1}{240}L \la +\frac{1}{1440}L' \la -\frac{1}{4320}L_{(2)}(\ga^{(2)}\la)
-\frac{31}{10080}L_{(4)}(\ga^{(4)}\la)\nn\\
&-\frac{1}{2880}K_{(1)}(\ga^{(1)}\mu)
-\frac{11}{11520}K_{(3)}(\ga^{(3)}\mu)
-\frac{1}{18432}(\la\ga_{(3)}\la)(\ga^{(3)}\mu)\nn\\
\widetilde{K}^{(3)}_{}&=-\frac{i}{360}\slsh D \mu
-\frac{1}{36}L' \la -\frac{1}{60}L_{(2)}(\ga^{(2)}\la)
+\frac{1}{180}L_{(4)}(\ga^{(4)}\la)\nn\\
&-\frac{2}{45}K_{(1)}(\ga^{(1)}\mu)
+\frac{1}{120}K_{(3)}(\ga^{(3)}\mu)
+\frac{1}{2880}(\la\ga_{(3)}\la)(\ga^{(3)}\mu)\nn\\
\widetilde{T}&=-\frac{8i}{45}\slsh D \mu
+\frac{8}{15}L\la+\frac{16}{45}L'\la
-\frac{8}{135}L_{(2)}(\ga^{(2)}\la)
-\frac{8}{45}L_{(4)}(\ga^{(4)}\la)\nn\\
&-\frac{56}{45}K_{(1)}(\ga^{(1)}\mu)
-\frac{8}{45}K_{(3)}(\ga^{(3)}\mu)
-\frac{1}{90}(\la\ga_{(3)}\la)(\ga_a^{(3)}\mu)
~.
\end{align}

{\it Vector-spinors}

Contracting the BI with the seven independent structures 
$(\ga^{eg_1g_2g_3})^{\ga}{}_{\bn}(\ga_{ag_1g_2g_3})_{\delta}{}^{\al}$, 
$(\ga_a{}^{eg_1g_2})^{\ga}{}_{\bn}(\ga_{g_1g_2})_{\delta}{}^{\al}$, 
$(\ga_{ag})^{\ga}{}_{\bn}(\ga^{eg})_{\delta}{}^{\al}$, 
$(\ga^{eg})^{\ga}{}_{\bn}(\ga_{ag})_{\delta}{}^{\al}$, 
$\delta_a^e\delta^{\ga}{}_{\bn}\delta_{\delta}{}^{\al}$, 
$(\ga_a{}^{eg_1\dots g_4})^{\ga}{}_{\bn}(\ga_{g_1\dots g_4})_{\delta}{}^{\al}$ and  
$(\ga_{ag_1g_2g_3})^{\ga}{}_{\bn}(\ga^{eg_1g_2g_3})_{\delta}{}^{\al}$, we obtain
\beal
\widetilde{L}^{(2){}}_{a}&=
-\frac{9i}{320}(\ga_a^{(1)} D_{(1)} \mu)-
\frac{21}{80}L_{(2)}(\ga_a^{(2)}\la)-
\frac{3}{40}L_{(4)}(\ga_a^{(4)}\la)\nn\\
&+\frac{3}{16}K_{(1)}(\ga_a^{(1)}\mu)
+\frac{63}{320}K_{(3)}(\ga_a^{(3)}\mu)
+\frac{9}{2560}(\la\ga_{(3)}\la)(\ga_a^{(3)}\mu)\nn\\
\widetilde{L}^{(4)}_{a{}}&=
-\frac{i}{1920}(\ga_a^{(1)} D_{(1)} \mu)-
\frac{1}{480}L_{(2)}(\ga_a^{(2)}\la)+
\frac{31}{1680}L_{(4)}(\ga_a^{(4)}\la)\nn\\
&-\frac{1}{480}K_{(1)}(\ga_a^{(1)}\mu)
+\frac{11}{4480}K_{(3)}(\ga_a^{(3)}\mu)
+\frac{1}{15360}(\la\ga_{(3)}\la)(\ga_a^{(3)}\mu)\nn\\
\widetilde{K}^{(3)}_{a}&=
-\frac{i}{160}(\ga_a^{(1)} D_{(1)} \mu)-\frac{1}{40}L_{(2)}(\ga_a^{(2)}\la)+
\frac{1}{20}L_{(4)}(\ga_a^{(4)}\la)\nn\\
&+\frac{1}{40}K_{(1)}(\ga_a^{(1)}\mu)
+\frac{81}{1120}K_{(3)}(\ga_a^{(3)}\mu)
+\frac{1}{1280}(\la\ga_{(3)}\la)(\ga_a^{(3)}\mu)\nn\\
\widetilde{T}_a&=-\frac{3i}{20}(\ga_a^{(1)} D_{(1)} \mu)-\frac{1}{5}L_{(2)}(\ga_{a}^{(2)}\la)
+\frac{2}{5}L_{(4)}(\ga_{a}^{(4)}\la)\nn\\
&+\frac{1}{5}K_{(1)}(\ga_{a}^{(1)}\mu)-\frac{3}{20}K_{(3)}(\ga_{a}^{(3)}\mu)
+\frac{3}{160}(\la\ga_{(3)}\la)(\ga_{a}^{(3)}\mu)~.
\end{align}

\vfill\break

{\it Two-form-spinors}

Contracting with the five independent structures 
$(\ga^e)_{\ga\bn}(\ga_{ab})_{\delta}{}^{\al}$, 
$(\ga_{ab}{}^{eg_1g_2})_{\ga\bn}(\ga_{g_1g_2})_{\delta}{}^{\al}$, 
$(\ga_{[a}{}^{eg})_{\ga\bn}(\ga_{b]g})_{\delta}{}^{\al}$, 
$(\ga_{g})_{\ga\bn}(\ga_{ab}{}^{eg})_{\delta}{}^{\al}$ and 
$(\ga^{eg_1g_2})_{\ga\bn}(\ga_{abg_1g_2})_{\delta}{}^{\al}$, we obtain
\beal
\widetilde{L}^{(2){}}_{ab}&=
\frac{3}{16}\widetilde{T}_{ab}
-\frac{1}{24}L_{(2)}(\ga_{ab}^{(2)}\la)-\frac{1}{4}L_{(4)}(\ga_{ab}^{(4)}\la)
-\frac{1}{8}K_{(3)}(\ga_{ab}^{(3)}\mu)\nn\\
\widetilde{L}^{(4)}_{ab{}}&=
\frac{1}{32}\widetilde{T}_{ab}
-\frac{1}{432}L_{(2)}(\ga_{ab}^{(2)}\la)
-\frac{13}{360}L_{(4)}(\ga_{ab}^{(4)}\la)
-\frac{1}{144}K_{(3)}(\ga_{ab}^{(3)}\mu)\nn\\
\widetilde{K}^{(3)}_{ab}&=
-\frac{1}{8}\widetilde{T}_{ab}-
\frac{1}{108}L_{(2)}(\ga_{ab}^{(2)}\la)+
\frac{1}{6}L_{(4)}(\ga_{ab}^{(4)}\la)
+\frac{1}{18}K_{(3)}(\ga_{ab}^{(3)}\mu)~.
\end{align}

{\it Three-form-spinors}

Contracting with the four independent structures 
$\delta^{\ga}{}_{\bn}(\ga_{abc}{}^{e})_{\delta}{}^{\al}$,  
$(\ga_{[a}{}^e)^{\ga}{}_{\bn}(\ga_{bc]})_{\delta}{}^{\al}$, 
$(\ga^{eg})^{\ga}{}_{\bn}(\ga_{abcg})_{\delta}{}^{\al}$ and 
$(\ga_{[a}{}^{eg_1g_2})^{\ga}{}_{\bn}(\ga_{bc]g_1g_2})_{\delta}{}^{\al}$, we obtain 
\beal
\widetilde{L}^{(4)}_{abc{}}&=
-\frac{1}{168}L_{(4)}(\ga_{abc}^{(4)}\la)
-\frac{1}{336}K_{(3)}(\ga_{abc}^{(3)}\mu)
\nn\\
\widetilde{K}^{(3)}_{abc}&=
-\frac{1}{84}K_{(3)}(\ga_{abc}^{(3)}\mu)~.
\end{align}

{\it Four-form-spinors}

Contracting with the independent structures $(\ga^e)_{\ga\bn}(\ga_{abcd})_{\delta}{}^{\al}$ and 
$(\ga_{[a}{}^{eg})_{\ga\bn}(\ga_{bcd]g})_{\delta}{}^{\al}$, we obtain 
\beal
\widetilde{L}^{(4)}_{abcd{}}&=0~.
\end{align}

$\bullet$  Case 2.2: $(\una,\ub, \und)=(\al, \upb, \upd)$

This is related to the previous case by parity-inversion. The analysis proceeds in a 
similar fashion.

{\it Spinors}

Contracting with the five independent structures 
$(\ga^e)^{\ga\al{}}\delta^{\delta}{}_{\bn}$, 
$(\ga_g)^{\ga\al{}}(\ga^{eg})^{\delta}{}_{\bn}$, 
$(\ga^{eg_1g_2})^{\ga\al{}}(\ga_{g_1g_2})^{\delta}{}_{\bn}$, \break
$(\ga_{g_1g_2g_3})^{\ga\al{}}(\ga^{eg_1g_2g_3})^{\delta}{}_{\bn}$ and 
$(\ga^{eg_1\dots g_4})^{\ga\al{}}(\ga_{g_1\dots g_4})^{\delta}{}_{\bn}$, 
we obtain
\beal
\widetilde{L}^{}&=\widetilde{L}^{\prime{}}-\frac{5i}{16}\slsh D \la+
\frac{1}{2}L \mu-2L' \mu -\frac{19}{6}L_{(2)}(\ga^{(2)}\mu)
+\frac{3}{2}L_{(4)}(\ga^{(4)}\mu)\nn\\
&-\frac{9}{4}K_{(1)}(\ga^{(1)}\la)
-\frac{9}{16}K_{(3)}(\ga^{(3)}\la)
-\frac{1}{128}(\mu\ga_{(3)}\mu)(\ga^{(3)}\la)\nn\\
\widetilde{L}^{(2){}}_{}&=\frac{13i}{960}\slsh D \la
-\frac{1}{120}L \mu -\frac{1}{5}L' \mu+\frac{47}{360}L_{(2)}(\ga^{(2)}\mu)
+\frac{1}{120}L_{(4)}(\ga^{(4)}\mu)\nn\\
&+\frac{5}{48}K_{(1)}(\ga^{(1)}\la)
-\frac{79}{960}K_{(3)}(\ga^{(3)}\la)
-\frac{7}{7680}(\mu\ga_{(3)}\mu)(\ga^{(3)}\la)\nn\\
\widetilde{L}^{(4)}_{{}}&=\frac{i}{11520}\slsh D \la
-\frac{1}{1440}L \mu +\frac{1}{240}L' \mu 
+\frac{1}{4320}L_{(2)}(\ga^{(2)}\mu)
-\frac{31}{10080}L_{(4)}(\ga^{(4)}\mu)
\nn\\
&+\frac{1}{2880}K_{(1)}(\ga^{(1)}\la)
-\frac{11}{11520}K_{(3)}(\ga^{(3)}\la)
+\frac{1}{18432}(\mu\ga_{(3)}\mu)(\ga^{(3)}\la)\nn\\
\widetilde{K}^{(3)}_{}&=\frac{i}{360}\slsh D \la
+\frac{1}{36}L \mu
 +\frac{1}{60}L_{(2)}(\ga^{(2)}\mu)
+\frac{1}{180}L_{(4)}(\ga^{(4)}\mu)\nn\\
&+\frac{2}{45}K_{(1)}(\ga^{(1)}\la)
+\frac{1}{120}K_{(3)}(\ga^{(3)}\la)
-\frac{1}{2880}(\mu\ga_{(3)}\mu)(\ga^{(3)}\la)\nn\\
\widetilde{T}&=-\frac{8i}{45}\slsh D \la
+\frac{16}{45}L\mu+\frac{8}{15}L'\mu
-\frac{8}{135}L_{(2)}(\ga^{(2)}\mu)
+\frac{8}{45}L_{(4)}(\ga^{(4)}\mu)\nn\\
&-\frac{56}{45}K_{(1)}(\ga^{(1)}\la)
+\frac{8}{45}K_{(3)}(\ga^{(3)}\la)
-\frac{1}{90}(\mu\ga_{(3)}\mu)(\ga_a^{(3)}\la)~.
\end{align}

{\it Vector-spinors}

Contracting the BI with the seven independent structures 
$(\ga^{eg_1g_2g_3})_{\ga}{}^{\al}(\ga_{ag_1g_2g_3})^{\delta}{}_{\bn}$, 
$(\ga_a{}^{eg_1g_2})_{\ga}{}^{\al}(\ga_{g_1g_2})^{\delta}{}_{\bn}$, 
$(\ga_{ag})_{\ga}{}^{\al}(\ga^{eg})^{\delta}{}_{\bn}$, 
$(\ga^{eg})_{\ga}{}^{\al}(\ga_{ag})^{\delta}{}_{\bn}$, 
$\delta_a^e\delta_{\ga}{}^{\al}\delta^{\delta}{}_{\bn}$, 
$(\ga_a{}^{eg_1\dots g_4})_{\ga}{}^{\al}(\ga_{g_1\dots g_4})^{\delta}{}_{\bn}$ and  
$(\ga_{ag_1g_2g_3})_{\ga}{}^{\al}(\ga^{eg_1g_2g_3})^{\delta}{}_{\bn}$, we obtain
\beal
\widetilde{L}^{(2){}}_{a}&=
-\frac{9i}{320}(\ga_a^{(1)} D_{(1)} \la)-
\frac{21}{80}L_{(2)}(\ga_a^{(2)}\mu)+
\frac{3}{40}L_{(4)}(\ga_a^{(4)}\mu)\nn\\
&+\frac{3}{16}K_{(1)}(\ga_a^{(1)}\la)
-\frac{63}{320}K_{(3)}(\ga_a^{(3)}\la)
+\frac{9}{2560}(\mu\ga_{(3)}\mu)(\ga_a^{(3)}\la)\nn\\
\widetilde{L}^{(4)}_{a{}}&=
\frac{i}{1920}(\ga_a^{(1)} D_{(1)} \la)+
\frac{1}{480}L_{(2)}(\ga_a^{(2)}\mu)+
\frac{31}{1680}L_{(4)}(\ga_a^{(4)}\mu)\nn\\
&+\frac{1}{480}K_{(1)}(\ga_a^{(1)}\la)
+\frac{11}{4480}K_{(3)}(\ga_a^{(3)}\la)
-\frac{1}{15360}(\mu\ga_{(3)}\mu)(\ga_a^{(3)}\la)\nn\\
\widetilde{K}^{(3)}_{a}&=
\frac{i}{160}(\ga_a^{(1)} D_{(1)} \la)+\frac{1}{40}L_{(2)}(\ga_{a}^{(2)}\mu)
+\frac{1}{20}L_{(4)}(\ga_{a}^{(4)}\mu)\nn\\
&-\frac{1}{40}K_{(1)}(\ga_{a}^{(1)}\la)+\frac{81}{1120}K_{(3)}(\ga_{a}^{(3)}\la)
-\frac{1}{1280}(\mu\ga_{(3)}\mu)(\ga_{a}^{(3)}\la)
\nn\\
\widetilde{T}_a&=-\frac{3i}{20}(\ga_a^{(1)} D_{(1)} \la)-\frac{1}{5}L_{(2)}(\ga_{a}^{(2)}\mu)
-\frac{2}{5}L_{(4)}(\ga_{a}^{(4)}\mu)\nn\\
&+\frac{1}{5}K_{(1)}(\ga_{a}^{(1)}\la)+\frac{3}{20}K_{(3)}(\ga_{a}^{(3)}\la)
+\frac{3}{160}(\mu\ga_{(3)}\mu)(\ga_{a}^{(3)}\la)~.
\end{align}

{\it Two-form-spinors}

Contracting with the five independent structures 
$(\ga^e)^{\ga\al{}}(\ga_{ab})^{\delta}{}_{\bn}$, 
$(\ga_{ab}{}^{eg_1g_2})^{\ga\al{}}(\ga_{g_1g_2})^{\delta}{}_{\bn}$, 
$(\ga_{[a}{}^{eg})^{\ga\al{}}(\ga_{b]g})^{\delta}{}_{\bn}$, 
$(\ga_{g})^{\ga\al{}}(\ga_{ab}{}^{eg})^{\delta}{}_{\bn}$ and 
$(\ga^{eg_1g_2})^{\ga\al{}}(\ga_{abg_1g_2})^{\delta}{}_{\bn}$, we obtain
\beal
\widetilde{L}^{(2){}}_{ab}&=
\frac{3}{16}\widetilde{T}_{ab}-\frac{1}{24}L_{(2)}(\ga_{ab}^{(2)}\mu)
+\frac{1}{4}L_{(4)}(\ga_{ab}^{(4)}\mu)
+\frac{1}{8}K_{(3)}(\ga_{ab}^{(3)}\la)\nn\\
\widetilde{L}^{(4)}_{ab{}}&=-
\frac{1}{32}\widetilde{T}_{ab}+\frac{1}{432}L_{(2)}(\ga_{ab}^{(2)}\mu)
-\frac{13}{360}L_{(4)}(\ga_{ab}^{(4)}\mu)-\frac{1}{144}K_{(3)}(\ga_{ab}^{(3)}\la)\nn\\
\widetilde{K}^{(3)}_{ab}&=
\frac{1}{8}\widetilde{T}_{ab}+\frac{1}{108}L_{(2)}(\ga_{ab}^{(2)}\mu)
+\frac{1}{6}L_{(4)}(\ga_{ab}^{(4)}\mu)+\frac{1}{18}K_{(3)}(\ga_{ab}^{(3)}\la)~.
\end{align}

{\it Three-form-spinors}

Contracting with the four independent structures 
$\delta_{\ga}{}^{\al}(\ga_{abc}{}^{e})^{\delta}{}_{\bn}$,  
$(\ga_{[a}{}^e)_{\ga}{}^{\al}(\ga_{bc]})^{\delta}{}_{\bn}$, 
$(\ga^{eg})_{\ga}{}^{\al}(\ga_{abcg})^{\delta}{}_{\bn}$ and 
$(\ga_{[a}{}^{eg_1g_2})_{\ga}{}^{\al}(\ga_{bc]g_1g_2})^{\delta}{}_{\bn}$, we obtain 
\beal
\widetilde{L}^{(4)}_{abc{}}&=-\frac{1}{168}L_{(4)}(\ga_{abc}^{(4)}\la)
-\frac{1}{336}K_{(3)}(\ga_{abc}^{(3)}\la)
\nn\\
\widetilde{K}^{(3)}_{abc}&=-\frac{1}{84}K_{(3)}(\ga_{abc}^{(3)}\la)~.
\end{align}

{\it Four-form-spinors}

Contracting with the independent structures $(\ga^e)^{\ga\al{}}(\ga_{abcd})^{\delta}{}_{\bn}$ and 
$(\ga_{[a}{}^{eg})^{\ga\al{}}(\ga_{bcd]g})^{\delta}{}_{\bn}$, we obtain 
\beal
\widetilde{L}^{(4)}_{abcd{}}&=0~.
\end{align}

$\bullet$ Case 3.1: $(\una,\ub, \und)=(\upa, \upb, \de)$

This is related to case 1.2 by parity-inversion.

{\it Spinors}

Contracting with the independent structures $(\ga^e)_{\al\bn}\delta^{\ga}{}_{\delta}$, 
$(\ga_g)_{\al\bn}(\ga^{ge})^{\ga}{}_{\delta}$ and 
$(\ga^{eg_1\dots g_4})_{\al\bn}(\ga_{g_1\dots g_4})^{\ga}{}_{\delta}$, we obtain
\beal
\widetilde{L}^{}&=\widetilde{L}^{\prime{}}+\frac{i}{16}\slsh D \la-
\frac{1}{4}L \mu-\frac{25}{8}L' \mu -\frac{73}{24}L_{(2)}(\ga^{(2)}\mu)
+\frac{9}{8}L_{(4)}(\ga^{(4)}\mu)\nn\\
&+\frac{3}{8}K_{(1)}(\ga^{(1)}\la)
-\frac{15}{16}K_{(3)}(\ga^{(3)}\la)
+\frac{1}{64}(\mu\ga_{(3)}\mu)(\ga^{(3)}\la)
+\frac{135}{64}\widetilde{T}
\nn\\
\widetilde{L}^{(2){}}_{}&=-\frac{3i}{320}\slsh D \la
+\frac{3}{80}L \mu -\frac{21}{160}L' \mu+\frac{59}{480}L_{(2)}(\ga^{(2)}\mu)
+\frac{1}{32}L_{(4)}(\ga^{(4)}\mu)\nn\\
&-\frac{9}{160}K_{(1)}(\ga^{(1)}\la)
-\frac{19}{320}K_{(3)}(\ga^{(3)}\la)
-\frac{3}{1280}(\mu\ga_{(3)}\mu)(\ga^{(3)}\la)
-\frac{33}{256}\widetilde{T}\nn\\
\widetilde{L}^{(4)}_{{}}&=\frac{i}{3840}\slsh D \la
-\frac{1}{960}L \mu +\frac{7}{1920}L' \mu 
+\frac{1}{3456}L_{(2)}(\ga^{(2)}\mu)
-\frac{131}{40320}L_{(4)}(\ga^{(4)}\mu)
\nn\\
&+\frac{1}{640}K_{(1)}(\ga^{(1)}\la)
-\frac{13}{11520}K_{(3)}(\ga^{(3)}\la)
+\frac{1}{15360}(\mu\ga_{(3)}\mu)(\ga^{(3)}\la)
+\frac{1}{1024}\widetilde{T}
~.
\end{align}

\vfill\break

{\it Vector-spinors}

Contracting with the three independent structures $\delta_a{}^e(\ga_g)_{\al\bn}(\ga^{g})_{\ga\delta}$, 
$(\ga_{ag_1\dots g_4})_{\al\bn}(\ga^{eg_1\dots g_4})_{\ga\delta}$ and \break
$(\ga_{a})_{\al\bn}(\ga^{e})_{\ga\delta}$, we obtain
\beal
\widetilde{L}^{(2){}}_{a}&=
-\frac{9i}{320}(\ga_a^{(1)} D_{(1)} \la)-
\frac{21}{80}L_{(2)}(\ga_a^{(2)}\mu)+
\frac{3}{40}L_{(4)}(\ga_a^{(4)}\mu)\nn\\
&+\frac{3}{16}K_{(1)}(\ga_a^{(1)}\la)
-\frac{63}{320}K_{(3)}(\ga_a^{(3)}\la)
+\frac{9}{2560}(\mu\ga_{(3)}\mu)(\ga_a^{(3)}\la)\nn\\
\widetilde{L}^{(4)}_{a{}}&=
\frac{i}{1920}(\ga_a^{(1)} D_{(1)} \la)+
\frac{1}{480}L_{(2)}(\ga_a^{(2)}\mu)+
\frac{31}{1680}L_{(4)}(\ga_a^{(4)}\mu)\nn\\
&+\frac{1}{480}K_{(1)}(\ga_a^{(1)}\la)
+\frac{11}{4480}K_{(3)}(\ga_a^{(3)}\la)
-\frac{1}{15360}(\mu\ga_{(3)}\mu)(\ga_a^{(3)}\la)\nn\\
\widetilde{T}_a&=-\frac{3i}{20}(\ga_a^{(1)} D_{(1)} \la)-\frac{1}{5}L_{(2)}(\ga_{a}^{(2)}\mu)
-\frac{2}{5}L_{(4)}(\ga_{a}^{(4)}\mu)\nn\\
&+\frac{1}{5}K_{(1)}(\ga_{a}^{(1)}\la)+\frac{3}{20}K_{(3)}(\ga_{a}^{(3)}\la)
+\frac{3}{160}(\mu\ga_{(3)}\mu)(\ga_{a}^{(3)}\la)
~.
\end{align}

{\it Two-form-spinors}

Contracting with the three independent structures 
$\delta_{[\al}{}^e(\ga_{b]})_{\al\bn}\delta^{\ga}{}_{\delta}$, 
$(\ga_{ab}{}^{eg_1g_2})_{\al\bn}(\ga_{g_1g_2})^{\ga}{}_{\delta}$, 
$\delta_{[\al}{}^e(\ga_{b]})_{\delta(\al}\delta^{\ga}_{\bn)}$
 and projecting onto the gamma-traceless part, we obtain
\beal
\widetilde{L}^{(2){}}_{ab}&=
\frac{3}{16}\widetilde{T}_{ab}-\frac{1}{24}L_{(2)}(\ga_{ab}^{(2)}\mu)
+\frac{1}{4}L_{(4)}(\ga_{ab}^{(4)}\mu)
+\frac{1}{8}K_{(3)}(\ga_{ab}^{(3)}\la)\nn\\
\widetilde{L}^{(4)}_{ab{}}&=-
\frac{1}{32}\widetilde{T}_{ab}+\frac{1}{432}L_{(2)}(\ga_{ab}^{(2)}\mu)
-\frac{13}{360}L_{(4)}(\ga_{ab}^{(4)}\mu)-\frac{1}{144}K_{(3)}(\ga_{ab}^{(3)}\la)
~.
\end{align}

{\it Three-form-spinors}

Contracting with $(\ga_{abc}{}^{eg})_{\al\bn}(\ga_{g})_{\ga\delta}$ we obtain
\beal
\widetilde{L}^{(4)}_{abc{}}&=-\frac{1}{168}L_{(4)}(\ga_{abc}^{(4)}\mu)
-\frac{1}{336}K_{(3)}(\ga_{abc}^{(3)}\la)
~.
\end{align}

{\it Four-form-spinors}

Contracting with the two independent structures 
$(\ga_{abcd}{}^{e})_{\al\bn}\delta^{\ga}{}_{\delta}$ and 
$(\ga_{abcd}{}^{e})_{\delta(\al}\delta^{\ga}{}_{\bn)}$, we obtain
\beal
\widetilde{L}^{(4)}_{abcd{}}&=0~.
\end{align}

$\bullet$  Case 3.2: $(\una,\ub, \und)=(\upa, \upb, \upd)$

This related to case 1.1 by parity inversion.

\vfill\break

{\it Spinors}

Contracting with the three independent structures $(\ga^e)_{\al\bn}\delta^{\delta}{}_{\ga}$, 
$(\ga_g)_{\al\bn}(\ga^{ge})^{\delta}{}_{\ga}$ and 
$(\ga^{eg_1\dots g_4})_{\al\bn}(\ga_{g_1\dots g_4})^{\delta}{}_{\ga}$, we obtain
\beal
\widetilde{T}&=-\frac{14i}{135}\slsh D \mu
+\frac{8}{45}L\la+\frac{32}{45}L'\la
+\frac{136}{405}L_{(2)}(\ga^{(2)}\la)
-\frac{8}{45}L_{(4)}(\ga^{(4)}\la)\nn\\
&-\frac{16}{45}K_{(1)}(\ga^{(1)}\mu)
+\frac{2}{45}K_{(3)}(\ga^{(3)}\mu)
-\frac{1}{60}(\la\ga_{(3)}\la)(\ga_a^{(3)}\mu)\nn\\
\widetilde{K}^{(3)}_{}&=\frac{i}{540}\slsh D \mu
-\frac{1}{45}L \la -\frac{1}{180}L' \la +\frac{13}{1620}L_{(2)}(\ga^{(2)}\la)
+\frac{1}{180}L_{(4)}(\ga^{(4)}\la)\nn\\
&+\frac{1}{90}K_{(1)}(\ga^{(1)}\mu)
+\frac{1}{45}K_{(3)}(\ga^{(3)}\mu)
~.
\end{align}

{\it Vector-spinors}

Contracting the BI with the four independent structures 
$\delta_a{}^e(\ga_g)_{\al\bn}(\ga^{g})_{\delta\ga}$,  
$\delta_a{}^e(\ga_{g_1\dots g_5})_{\al\bn}(\ga^{g_1\dots g_5})_{\delta\ga}$, \break 
$(\ga_a)_{\al\bn}(\ga^{e})_{\delta\ga}$, 
$(\ga_{a g_1\dots g_4})_{\al\bn}(\ga^{eg_1\dots g_4})_{\delta\ga}$ and 
projecting onto the gamma-traceless part, we obtain
\beal
\widetilde{T}_a&=-\frac{3i}{20}(\ga_a^{(1)} D_{(1)} \mu)-\frac{1}{5}L_{(2)}(\ga_{a}^{(2)}\la)
+\frac{2}{5}L_{(4)}(\ga_{a}^{(4)}\la)\nn\\
&+\frac{1}{5}K_{(1)}(\ga_{a}^{(1)}\mu)-\frac{3}{20}K_{(3)}(\ga_{a}^{(3)}\mu)
+\frac{3}{160}(\la\ga_{(3)}\la)(\ga_{a}^{(3)}\mu)\nn\\
\widetilde{K}^{(3)}_{a}&=
-\frac{i}{160}(\ga_a^{(1)} D_{(1)} \mu)-\frac{1}{40}L_{(2)}(\ga_a^{(2)}\la)+
\frac{1}{20}L_{(4)}(\ga_a^{(4)}\la)\nn\\
&+\frac{1}{40}K_{(1)}(\ga_a^{(1)}\mu)
+\frac{81}{1120}K_{(3)}(\ga_a^{(3)}\mu)
+\frac{1}{1280}(\la\ga_{(3)}\la)(\ga_a^{(3)}\mu)
~.
\end{align}
The notation in the equation above is a shorthand for the projection onto the 
irreducible (gamma-traceless) vector-spinor part. Our conventions are 
explained in appendix \ref{gammatraceless}.

{\it Two-form-spinors}

Contracting with the two independent structures 
$(\ga_{[a})_{\al\bn}(\ga_{b]}{}^{e})^{\delta}{}_{\ga}$, 
$(\ga_{ab}{}^{eg_1g_2})_{\al\bn}(\ga_{g_1g_2})^{\delta}{}_{\ga}$, we obtain
\beal
\widetilde{K}^{(3)}_{ab}&=
-\frac{1}{8}\widetilde{T}_{ab}-
\frac{1}{108}L_{(2)}(\ga_{ab}^{(2)}\la)+
\frac{1}{6}L_{(4)}(\ga_{ab}^{(4)}\la)
+\frac{1}{18}K_{(3)}(\ga_{ab}^{(3)}\mu)
~.
\end{align}
As in the previous case, the projections onto the gamma-traceless part of the 
two-form-spinors are explained in appendix \ref{gammatraceless}.

{\it Three-form-spinors}

Contracting with the two independent structures 
$\delta_{[a}{}^e(\ga_{bc]g_1g_2g_3})_{\al\bn}(\ga^{g_1g_2g_3})_{\delta\ga}$ 
and $(\ga_{abc}{}^{eg})_{\al\bn}(\ga_{g})_{\delta\ga}$, we obtain
\beal
\widetilde{K}^{(3)}_{abc}&=
-\frac{1}{84}K_{(3)}(\ga_{abc}^{(3)}\mu)
~.
\end{align}

{\it Four-form-spinors}

Contracting with 
$(\ga_{abcd}{}^e)_{\al\bn}\delta^{\delta}{}_{\ga}$, we obtain
\beal
\widetilde{L}^{(4)}_{abcd{}}&=0~.
\end{align}
The solution of the (highly overdetermined) system of equations above is given in section 
\ref{summary}. Note that the spinor equations for $(\widetilde{L}-\widetilde{L}')^{\al}$ and 
$(\widetilde{L}-\widetilde{L}')_{\al}$ (equations 
(\ref{rightspinorl},\ref{leftspinorl}), respectively) can 
actually be solved for $(L-L')$, as we now show. Let us  
parameterize $(L-L')$ as follows
\beal
L-L'=\frac{3}{2}(\mu\la)+me^{2\phi}
~,
\label{ktulu}
\end{align}
where $m$ is a massive constant and $\phi$ is some real scalar superfield 
(possibly a constant) of canonical 
mass dimension zero. Plugging (\ref{ktulu}) in (\ref{rightspinorl},\ref{leftspinorl}) 
and taking into account the action of the spinor derivative on $\la_{\al}, \mu^{\al}$,  
we obtain
\beal
m({D}_{\al}\phi-\la_{\al})=0
\end{align}
and 
\beal
m({D}^{\al}\phi-\mu^{\al})=0~.
\end{align}
These equations have two possible solutions: (a) $m=0$ or 
(b) $m\neq 0$ and ${D}^{\al}\phi=\mu^{\al}$, ${D}_{\al}\phi=\la_{\al}$. 
We shall see in section \ref{dimension2} that case (b) is that of Romans supergravity. 
Moreover let us distinguish two further subcases: (a1) $m=0$, $L=-L'$ and 
(a2)  $m=0$,  $L\neq -L'$. 
As we show in the following,  case (a1) is that of  massless IIA 
while (a2) is that of HLW supergravity.

Finally, the spinor derivatives of $K_a$ are computed by employing the identities 
\beal
2D_{(\al}D_{\bn)}\la_{\ga}&=-T_{\al\bn}{}^ED_{E}\la_{\ga}-R_{\al\bn\ga}{}^{\delta}\la_{\delta}\nn\\
2D^{(\al}D^{\bn)}\mu^{\ga}&=-T^{\al\bn E}D_{E}\mu^{\ga}+\mu^{\delta}R^{\al\bn}{}_{\delta}{}^{\ga}
~.
\end{align}
The result is given in section \ref{summary}.

\subsection{Dimension-2 BI}
\label{dimension2}

Taking into account our 
gauge-fixing  for the supertorsion components, the first BI at dimension two reads
\beal
R_{[abc]}{}^d=0~.
\end{align}
The second BI at dimension two can be cast in the form
\beal
D_{\una}\widetilde{T}_{ab}{}^{\und}=R_{ab\una}{}^{\und}-2D_{[a}T_{b]\una}{}^{\und}
-T_{ab}{}^{\une}T_{\une\una}{}^{\und}
-2T_{\una[a|}{}^{\une}T_{\une|b]}{}^{\und}-D_{\una}(\ga_{[a}\widetilde{T}_{b]})^{\und}
-D_{\una}(\ga_{ab}\widetilde{T}_{})^{\und}
~.
\label{poui}
\end{align}
This simply determines the spinor derivative of  $\widetilde{T}_{ab}$, 
provided the right-hand-side of the equation above 
is consistent with the gamma-tracelessness of $\widetilde{T}_{ab}$. 
In other words, the right-hand-side of (\ref{poui})  
should be annihilated by $(\ga_a)^{\uc}{}_{\und}$. This condition 
 turns out to be equivalent to the system of 
all equations-of-motion and Bianchi identities 
for the bosonic fields! To see this in detail, let us 
distinguish the following cases. 

\vfill\break

$\bullet$  Case 1: $(\una, \und)=(\al, \de)$

For simplicity of presentation, here and in the remainder of this section 
we shall focus on the bosonic terms, ignoring quadratic and quartic fermionic terms. 
In this case, the condition for the 
tracelessness of the right-hand-side of (\ref{poui}) can be written as
\beal
0=\ga^c\mathcal{A}_{bc}+\ga^{cde}\mathcal{A}_{bcde}+\ga^{cdefg}\mathcal{A}_{bcdefg}
~,
\end{align}
where the coefficients in the expansion above are given by
\beal
\mathcal{A}_{bc}&=
\eta_{bc}\Big(
-\frac{72}{25}L L'-\frac{3i}{5}D^iK_i-\frac{36}{5}K^iK_i
-\frac{304}{45}L_{ij}L^{ij}-\frac{192}{5}K_{ijk}K^{ijk}-\frac{576}{5}L_{ijkl}L^{ijkl}
\Big)\nn\\
&+\frac{27i}{5}D_{b}K_c+\frac{3i}{5}D_cK_b-8K_bK_c
-\frac{144}{25}LL_{bc}-\frac{176}{25}L'L_{bc}-\frac{64}{9}L_b{}^iL_{ci}
+\frac{18i}{5}D^iK_{ibc}
\nn\\
&+\frac{144}{5}K^iK_{ibc}+72K_b{}^{ij}K_{cij}
+\frac{384}{5}L^{ij}L_{ijbc}-1536L_b{}^{ijk}L_{cijk}
+\frac{24}{5}\varepsilon_{bci_1\dots i_8}L^{i_1\dots i_4}L^{i_5\dots i_8}
-\frac{1}{2}R_{bc}
\nn\\
\mathcal{A}_{bcde}&=
\eta_{b[c}\Big(
\frac{3i}{5}D_dK_{e]}-\frac{24}{5}K^iK_{i|de]}
-\frac{192}{5}K^iK_{i|de]}-\frac{124}{75}L L_{|de]}
+\frac{4}{75}L'L_{|de]}\nn\\
&-\frac{512}{5}L^{ij}L_{ij|de]}
-\frac{32}{5}\varepsilon_{|de]i_1\dots i_8}L^{i_1\dots i_4}L^{i_5\dots i_8}\Big)
-\frac{1}{4}R_{b[cde]}\nn\\
&-\frac{12i}{5}D_{[b}K_{cde]}-\frac{48}{5}K_{[b}K_{cde]}
+\frac{48}{5}LL_{bcde}+\frac{48}{5}L'L_{bcde}
\nn\\
\mathcal{A}_{bcdefg}&=\eta_{b[c}\Big(
\frac{8i}{5}D_{d}K_{efg]}+\frac{32}{5}K_{|d}K_{efg]}-\frac{32}{5}LL_{|defg]}
-\frac{32}{5}L'L_{|defg]}
\Big)
~.
\end{align}
To arrive at these expressions we have Hodge-dualized the $\ga^{(7)}$ and $\ga^{(9)}$ terms. 
We have also made use of the identity
\beal
\varepsilon_{bci_1\dots i_8}L^{i_1\dots i_4}L^{i_5\dots i_8}
(\ga_a{}^{bc})_{\al\bn}=16(\ga^{i_1\dots i_7})_{\al\bn}
L_{ai_1i_2i_3}L_{i_4\dots i_7}~.
\end{align}

$\bullet$  Case 2: $(\una, \und)=(\al, \upd)$

In this case the tracelessness condition takes the form
\beal
0=\mathcal{B}_b
+\ga^{cd}\mathcal{B}_{bcd}+\ga^{cdef}\mathcal{B}_{bcdef}~,
\end{align}
where 
\beal
\mathcal{B}_b&=
\frac{9i}{5}D_bL+\frac{18i}{5}D_bL'-8 LK_b-\frac{76}{5}L'K_b
+\frac{58i}{15}D^iL_{ib}+\frac{232}{5}K^iL_{ib}
+\frac{2784}{5}K^{ijk}L_{ijkb}
\nn\\
\mathcal{B}_{bcd}&=
\eta_{b[c}\Big(
iD_{d]}L-\frac{2i}{5}D_{d]}L'+\frac{24}{5}LK_{|d]}+\frac{52}{5}L'K_{|d]}
-\frac{38i}{15}D^iL_{i|d]}-\frac{152}{5}K^iL_{i|d]}
-\frac{1824}{5}K^{ijk}L_{ijk|d]}
\Big)\nn\\
&-\frac{216}{25}LK_{bcd}-\frac{564}{25}L'K_{bcd}
-\frac{29i}{5}D_{[b}L_{cd]}+\frac{156}{5}D^iL_{ibcd}
+\frac{1248}{5}K^iL_{ibcd}
-\frac{26}{5}\varepsilon_{bcdi_1\dots i_7}K^{i_1i_2i_3}L^{i_4\dots i_7}
\nn\\
\mathcal{B}_{bcdef}&=
\eta_{b[c}\Big(
-\frac{8}{25}LK_{def}+\frac{68}{25}L'K_{def}
+\frac{19i}{5}D_{|d}L_{ef]}\nn\\
&-\frac{12i}{5}D^iL_{i|def]}
-\frac{96}{5}K^iL_{i|def]}
+\frac{2}{5}\varepsilon_{|def]i_1\dots i_7}K^{i_1i_2i_3}L^{i_4\dots i_7}\Big)\nn\\
&+\frac{52}{3}\Big( 
L_{[bc}K_{def]}
-\frac{3}{4}iD_{[b}L_{cdef]}
-3K_{[b}L_{cdef]}
\Big)\nn\\
&-\frac{1}{30}\varepsilon_{bcdefi_1\dots i_5}\Big(
L^{i_1i_2}K^{i_3i_4i_5}
-\frac{3i}{4}D^{i_1}L^{i_2\dots i_4}
-3K^{i_1}L^{i_2\dots i_4}
\Big)
~.
\end{align}
As in the previous case, we have Hodge-dualized the $\ga^{(6)}$, $\ga^{(8)}$, $\ga^{(10)}$ 
contributions. Also, we have taken into 
account the identity
\beal
\frac{1}{24}\varepsilon_{cdei_1\dots i_7}(\ga_b{}^{cde})_{\al}{}^{\delta}K^{i_1i_2i_3}L^{i_4\dots i_7}
=-(\ga_{i_1\dots i_6})_{\al}{}^{\delta}\Big(L_b{}^{i_1i_2i_3}K^{i_4i_5i_6}+\frac{3}{4}
K_b{}^{i_1i_2}L^{i_3\dots i_6}
\Big)~.
\end{align}

$\bullet$  Case 3: $(\una, \und)=(\upa, \de)$

This is similar to the previous case. The gamma-tracelessness condition is 
of the form 
\beal
0=\mathcal{C}_b
+\ga^{cd}\mathcal{C}_{bcd}+\ga^{cdef}\mathcal{C}_{bcdef}~.
\end{align}
The coefficients $\mathcal{C}_b$, 
$\mathcal{C}_{bcd}$ and $\mathcal{C}_{bcdef}$ can be obtained from 
$\mathcal{B}_b$, 
$\mathcal{B}_{bcd}$ and $\mathcal{B}_{bcdef}$ respectively, 
by making the substitutions $L\leftrightarrow L'$, 
$K_{(3)}\rightarrow -K_{(3)}$, $L_{4}\rightarrow -L_{4}$ and 
$\varepsilon_{(10)}\rightarrow -\varepsilon_{(10)}$.

$\bullet$  Case 4: $(\una, \und)=(\upa, \upd)$

This is related to case 1 by parity inversion. The gamma-tracelessness 
condition takes the form 
\beal
0=\ga^c\mathcal{D}_{bc}+\ga^{cde}\mathcal{D}_{bcde}+\ga^{cdefg}\mathcal{D}_{bcdefg}
~.
\end{align}
The coefficients $\mathcal{D}_{bc}$, $\mathcal{D}_{bcde}$ and 
$\mathcal{D}_{bcdefg}$ above can be obtained from 
$\mathcal{A}_{bc}$, $\mathcal{A}_{bcde}$ and 
$\mathcal{A}_{bcdefg}$ respectively, by making the 
substitutions described in the previous case.

In conclusion, the gamma-tracelessness condition is equivalent to 
\beal
\mathcal{A}=\mathcal{B}=\mathcal{C}=\mathcal{D}=0~.
\label{gt}
\end{align}
It is straightforward to recognize that in the case (b) 
at the end of 
the previous subsection   
(i.e. for $L-L'=\frac{3}{2}(\mu\la)+me^{2\phi}$, $L+L'=0$ and 
$\la_{\al}=D_{\al}\phi$, $\mu^{\al}=D^{\al}\phi$, 
equations (\ref{gt}) reduce to those of Romans supergravity. 

In case (a1) (i.e. for $L-L'=\frac{3}{2}(\mu\la)$, $L+L'=0$) 
we can see, using the results of the preceding analysis 
of the BI's, that 
the super-one-form $\Lambda_A$ defined by
\beal
\Lambda_{\al}&:=\la_{\al}\nn\\
\Lambda^{\al}&:=\mu^{\al}\nn\\
\Lambda_{a}&:=-2iK_{a}
~,
\end{align}
is closed:
\beal
D\Lambda=0~.
\label{clos}
\end{align}
It follows that there exists a scalar superfield $\phi$ 
such that $\Lambda=D\phi$. In particular, 
$\la_{\al}=D_{\al}\phi$, $\mu^{\al}=D^{\al}\phi$ and $K_a=\frac{i}{2}D_a\phi$. It is now 
straightforward to see that in this case equations (\ref{gt}) reduce 
to those of massless type IIA supergravity \cite{iia}. Note that (\ref{clos}) 
also holds in the case of Romans supergravity.

In the case where spacetime $\mathcal{M}$ is not simply connected, 
it may be that $\Lambda=D\phi+\psi$, where $\psi$ is 
closed but not exact. For example, if $\mathcal{M}$ 
contains an $S^1$ we may take $\psi=m dz$, where 
$z$ is the $S^1$ coordinate and $m$ is a mass parameter. 
Upon compactification on $S^1$, this would amount to 
a Scherk-Schwarz reduction. The possibility for this 
topological modification is the remnant in ten dimensions of 
the freedom to modify ordinary eleven-dimensional 
supergravity to MM-theory.

Finally, in case (a2)  
($L-L'=\frac{3}{2}(\mu\la)$, $L+L'\neq 0$), equations 
(\ref{gt}) reduce to those of HLW supergravity,  
 summarized in section \ref{summary}. 
As we have already remarked at the end of section \ref{dimension1}, 
in this case there cannot exist a $\phi$ such that 
$D_{\al}\phi=\la_{\al}$, $D^{\al}\phi=\mu^{\al}$. 
Note that if we parameterize
\beal
L&=\frac{3}{2}\Big(\Phi+\frac{1}{2}(\mu\la)\Big)\nn\\
L'&=\frac{3}{2}\Big(\Phi-\frac{1}{2}(\mu\la)\Big)~,
\end{align}
where the massless IIA limit (case (a1) above) is reached at $\Phi\rightarrow 0$, 
equations (\ref{gt}) imply in particular that $\Phi=$constant. 
In equations (\ref{hl1}, \ref{hl2}) of section \ref{summary} we have set $\Phi=m$.

\subsection{Dimension-$\frac{5}{2}$ BI}

It can be seen that the dimension-$\frac{5}{2}$ BI does not introduce any new constraints, 
other than determining the action of the spinor derivative 
on the dimension-2 component of the supercurvature. Explicitly:
\beal
D_{\una}R_{abcd}
=2D_{[a|}R_{\una|b]cd}-T_{ab}{}^{\une} R_{\une \una cd}
+2T_{[a|\una}{}^{\une} R_{\une|b]cd} ~.
\end{align}

\section{Summary}
\label{summary}

Here we summarize the result of the 
analysis of the BI's carried out in the preceding section. 
At each order in canonical mass dimension we give the solution for the components 
of the torsion and curvature, as well as the action of the spinor 
derivative on the various superfields at that dimension\footnote{
We have checked that our formul{\ae} 
are compatible with overlapping literature. 
For example, in order to compare with \cite{cede}, 
where the supertorsion components 
were given up to dimension-$\frac{1}{2}$, one should make the 
following identifications: $\Lambda_{\al}\rightarrow (\mu^{\al}, \la_{\al})$ and 
$$
(\ga_{11})_{\al}{}^{\bn}
\rightarrow \left(\begin{array}{cc}
-\de_{\al}{}^{\bn} & 0\\
0& \de^{\al}{}_{\bn}
\end{array}\right)~.
$$
One also needs to take into account the 
identity 
$$
(\ga^{ab})_{(\al}{}^{\ga}(\ga_{ab}\la)_{\bn)}
=-10\de_{(\al}^{\ga}\la_{\bn)}+4\ga^a_{\al\bn}(\ga_a\la)^{\ga}~.
$$
}. At dimension two we have chosen to include the 
bosonic part of the equations-of-motion and Bianchi identities for the 
bosonic fields of the relatively less-known HLW supergravity.

\subsection*{Dimension $0$}
\beal
T_{\al \bn}{}^c&=-i(\ga^c)_{\al\bn}\nn\\
T^{\al \bn c}&=-i(\ga^c)^{\al\bn}\nn\\
T_\al{}^{\bn c}&=0
~.
\label{tz}
\end{align}

\subsection*{Dimension $\frac{1}{2}$}
\beal
T_{\al b}{}^c&=T^{\al}{}_{b}{}^c=0\nn\\
T_{\al\bn}{}^\ga&=-4\Big\{
\la_{(\al}
\de_{\bn)}{}^{\ga}-\frac{1}{2}(\ga^e)_{\al\bn}(\ga_e\la)^{\ga} \Big\}  \nn\\
T^{\al\bn}{}_\ga&=-4\Big\{
\mu^{(\al}
\de^{\bn)}{}_{\ga}-\frac{1}{2}(\ga_e)^{\al\bn}(\ga^e\mu)_{\ga}\Big\} \nn\\
T_{\al\bn\ga}&=(\ga^e)_{\al\bn}(\ga_e\mu)_\ga\nn\\
T_\al{}^{\bn\ga}&=-\frac{1}{2}\Big\{
\de_\al{}^\ga\mu^\bn -\frac{1}{2}(\ga^{ef})_\al{}^\ga(\ga_{ef}\mu)^\bn
\Big\}\nn\\
T^\al{}_{\bn\ga}&=-\frac{1}{2} \Big\{
\de^\al{}_\ga\la_\bn -\frac{1}{2}(\ga^{ef})^\al{}_\ga(\ga_{ef}\la)_\bn
\Big\}\nn\\
T^{\al\bn\ga}&=(\ga_e)^{\al\bn}(\ga^e\la)^\ga~.
\end{align}
%

\subsection*{Dimension $1$}

Torsion:
\beal
T_{ab}{}^c&=0\nn\\
T_{a\al}{}^\bn&= H^1_{fgh}
(\ga_a{}^{fgh})_\al{}^\bn +H^2_{afg}(\ga^{fg})_\al{}^\bn 
\nn\\
T_a{}^{\al}{}_{\bn}&= 
H^{\prime 1}_{fgh}(\ga_a{}^{fgh})^\al{}_\bn 
+H^{\prime 2}_{afg}(\ga^{fg})^\al{}_\bn 
\nn\\
T_{a\al\bn}&=S 
(\ga_a)_{\al\bn} +F^1_{ef}(\ga_a{}^{ef})_{\al\bn} 
+F^2_{ae}(\ga^{e})_{\al\bn}
+G^1_{efgh} (\ga_a{}^{efgh})_{\al\bn}
+G^2_{aefg}(\ga^{efg})_{\al\bn}
\nn\\
T_a{}^{\al\bn}&=
-S(\ga_a)^{\al\bn} +F^1_{ef}(\ga_a{}^{ef})^{\al\bn} 
+F^2_{ae}(\ga^{e})^{\al\bn}
-G^1_{efgh} (\ga_a{}^{efgh})^{\al\bn}
-G^2_{aefg}(\ga^{efg})^{\al\bn}
~,
\end{align}
Curvature:
\beal
R_{\al\bn} {}_{cd}&=2iH^1_{efg}(\ga_{cd}{}^{efg})_{\al\bn}
+4iH^2_{cde}(\ga^{e})_{\al\bn}\nn\\
R^{\al\bn} {}_{cd}&=2iH^{\prime 1}_{efg}(\ga_{cd}{}^{efg})^{\al\bn}
+4iH^{\prime 2}_{cde}(\ga^{e})^{\al\bn}\nn\\
R_{\al}{}^{\bn} {}_{cd}&=-2i\Big\{S
(\ga_{cd})_{\al}{}^\bn +F^1_{ef}(\ga_{cd}{}^{ef})_{\al}{}^\bn 
+F^2_{cd} \de_{\al}{}^\bn +G^1_{efgh}
(\ga_{cd}{}^{efgh})_{\al}{}^\bn 
+3G^2_{cdfg}(\ga^{fg})_{\al}{}^\bn
\Big\}~.
\end{align}
Spinor derivatives:
\beal
D_{\al}\la_\bn&=K_e(\ga^e)_{\al\bn}+K_{efg}(\ga^{efg})_{\al\bn}\nn\\
D_\al\mu^\bn&=L~\de_\al{}^\bn+L_{ef}(\ga^{ef})_\al{}^\bn
+L_{efgh}(\ga^{efgh})_\al{}^\bn\nn\\
D^{\al}\la_\bn&={L'}~\de^\al{}_\bn+L^{ef}(\ga_{ef})^\al{}_\bn
-L^{efgh}(\ga_{efgh})^\al{}_\bn\nn\\
D^\al\mu^\bn&={K}^{e}(\ga_e)^{\al\bn}
-{K}^{efg}(\ga_{efg})^{\al\bn}~,
\end{align}
where
\beal
S&=
-\frac{2i}{5}(L-{L'})+\frac{3i}{5}(\mu\la)\nn\\
F_{ab}^1&=-\frac{2i}{3}L_{ab}\nn\\
F_{ab}^2&=\frac{8i}{3}L_{ab}+i(\mu\ga_{ab}\la)\nn\\
H_{abc}^1&=iK_{abc}\nn\\
H_{abc}^2&=-\frac{i}{8}(\la\ga_{abc}\la)-\frac{i}{8}(\mu\ga_{abc}\mu)\nn\\
{H}_{abc}^{\prime 1}&=-iK_{abc}\nn\\
{H}_{abc}^{\prime 2}&=-\frac{i}{8}(\la\ga_{abc}\la)-\frac{i}{8}(\mu\ga_{abc}\mu)\nn\\
G^1_{abcd}&=0\nn\\
G^2_{abcd}&=8iL_{abcd}~.
\end{align}
For the scalar fields $L$, $L'$, we have the following 
cases\footnote{In fact, massless IIA is the massless limit of both 
Romans and HLW supergravity and need not be presented as a separate case.}
\beal
\mbox{Massless IIA:}&  \left\{
\begin{array}{l}
L=\frac{3}{4}(\mu\la)\\
L'=-\frac{3}{4}(\mu\la)
\end{array}\right.\nn\\
\mbox{Romans:}&\left\{
\begin{array}{l}
L=\frac{1}{2}me^{2\phi}+\frac{3}{4}(\mu\la)\\
L'=-\frac{1}{2}me^{2\phi}-\frac{3}{4}(\mu\la)
\end{array}\right.\nn\\
\mbox{HLW:}& \left\{
\begin{array}{l}
L=\frac{3}{2}m+\frac{3}{4}(\mu\la)\\
L'=\frac{3}{2}m-\frac{3}{4}(\mu\la)
\end{array}\right.
~,
\label{llprime}
\end{align}
where in both Romans and massless IIA supergravities we have 
(modulo the possibility for the topological modification 
of massless IIA explained at the end of section \ref{dimension2}) 
$D_{\al}\phi=\la_{\al}$, 
$D^{\al}\phi=\mu^{\al}$. In HLW supergravity, $\nexists$ $\phi$:  
$D_{\al}\phi=\la_{\al}$, $D^{\al}\phi=\mu^{\al}$.

\subsection*{Dimension $\frac{3}{2}$}

{\it Torsion}
\beal
T_{ab}=\widetilde{T}_{ab}
+{}\ga_{[a}\widetilde{T}_{b]}{}{}
+{}\ga_{ab}\widetilde{T}
~,
\end{align}
where for the right-handed spinors $\widetilde{T}^{\al}$, $\widetilde{T}_a^{\al}$ we have 
(suppressing spinor indices)
\beal
\widetilde{T}&=\frac{272}{225}L\mu-\frac{8}{25}L'\mu
+\frac{8}{9}L_{(2)}(\ga^{(2)}\mu)
+\frac{8}{45}L_{(4)}(\ga^{(4)}\mu)\nn\\
&+\frac{8}{9}K_{(1)}(\ga^{(1)}\la)
-\frac{16}{45}K_{(3)}(\ga^{(3)}\la)
-\frac{11}{450}(\mu\ga_{(3)}\mu)(\ga^{(3)}\la)
\nn\\
\widetilde{T}_a&=-\frac{3i}{20}(\ga_a^{(1)} D_{(1)} \mu)-\frac{1}{5}L_{(2)}(\ga_{a}^{(2)}\la)
+\frac{2}{5}L_{(4)}(\ga_{a}^{(4)}\la)\nn\\
&+\frac{1}{5}K_{(1)}(\ga_{a}^{(1)}\mu)-\frac{3}{20}K_{(3)}(\ga_{a}^{(3)}\mu)
+\frac{3}{160}(\la\ga_{(3)}\la)(\ga_{a}^{(3)}\mu)
\label{mi}
\end{align}
and similarly for the left-handed spinors $\widetilde{T}_{\al}$, $\widetilde{T}_{a\al}$,
\beal
\widetilde{T}&=
-\frac{8}{25}L\la+\frac{272}{225}L'\la
+\frac{8}{9}L_{(2)}(\ga^{(2)}\la)
-\frac{8}{45}L_{(4)}(\ga^{(4)}\la)\nn\\
&+\frac{8}{9}K_{(1)}(\ga^{(1)}\mu)
+\frac{16}{45}K_{(3)}(\ga^{(3)}\mu)
-\frac{11}{450}(\la\ga_{(3)}\la)(\ga^{(3)}\mu)\nn\\
\widetilde{T}_a&=-\frac{3i}{20}(\ga_a^{(1)} D_{(1)} \la)-\frac{1}{5}L_{(2)}(\ga_{a}^{(2)}\mu)
-\frac{2}{5}L_{(4)}(\ga_{a}^{(4)}\mu)\nn\\
&+\frac{1}{5}K_{(1)}(\ga_{a}^{(1)}\la)+\frac{3}{20}K_{(3)}(\ga_{a}^{(3)}\la)
+\frac{3}{160}(\mu\ga_{(3)}\mu)(\ga_{a}^{(3)}\la)
~.
\label{ni}
\end{align}
No confusion should arise from the slight abuse of notation, as it is 
immediately clear by the right-hand-sides of equations (\ref{mi},\ref{ni})
what the chiralities of the spinors $\widetilde{T}$ are in each case.

{\it Curvature}
\beal
R_{\al bcd}&=\frac{i}{2}\big(
\ga_b T_{cd} +\ga_c T_{bd}-\ga_d T_{bc}
\big)_{\al}\nn\\
R^{\al}{}_{ bcd}&=\frac{i}{2}\big(
\ga_b T_{cd} +\ga_c T_{bd}-\ga_d T_{bc}
\big)^{\al}
~.
\end{align}

{\it Spinor derivatives}
\beal
D{}L&=\widetilde{L}{}\nn\\
D{}L'&=\widetilde{L}^{\prime}\nn\\
D{}L_{ab}&=\widetilde{L}^{(2)}_{ab\al}
+{}\ga_{[a}\widetilde{L}^{(2)}_{b]}{}{}
+{}\ga_{ab}\widetilde{L}^{(2)}{}{}\nn\\
D{}L_{abcd}&=\widetilde{L}^{(4)}_{abcd\al}
+{}\ga_{[a}\widetilde{L}^{(4)}_{bcd]}{}{}
+{}\ga_{[ab}\widetilde{L}^{(4)}_{cd]}{}{}
+{}\ga_{abc}\widetilde{L}^{(4)}_{d]}{}{}
+{}\ga_{abc}\widetilde{L}^{(4)}{}{}
~,
\end{align}
\beal
D{}K_{a}&=\widetilde{K}^{(1)}_{a\al}
+{}\ga_{a}\widetilde{K}^{(1)}{}{}\nn\\
D{}K_{abc}&=\widetilde{K}^{(3)}_{abc\al}
+{}\ga_{[a}\widetilde{K}^{(3)}_{bc]}{}{}
+{}\ga_{[ab}\widetilde{K}^{(3)}_{c]}{}{}
+{}\ga_{abc}\widetilde{K}^{(3)}{}{}
~,
\end{align}
where
\beal
\widetilde{L}^{}&=\widetilde{L}^{\prime{}}+
2L \mu-\frac{7}{2}L' \mu -\frac{3}{2}L_{(2)}(\ga^{(2)}\mu)
+\frac{3}{2}L_{(4)}(\ga^{(4)}\mu)\nn\\
&+\frac{3}{2}K_{(1)}(\ga^{(1)}\la)
-\frac{3}{2}K_{(3)}(\ga^{(3)}\la)
-\frac{1}{32}(\mu\ga_{(3)}\mu)(\ga^{(3)}\la)~,
\label{rightspinorl}
\end{align}
\beal
\widetilde{L}^{(2){}}_{ab}&=
\frac{3}{16}\widetilde{T}_{ab}-\frac{1}{24}L_{(2)}(\ga_{ab}^{(2)}\mu)
+\frac{1}{4}L_{(4)}(\ga_{ab}^{(4)}\mu)
+\frac{1}{8}K_{(3)}(\ga_{ab}^{(3)}\la)\nn\\
\widetilde{L}^{(2){}}_{a}&=
-\frac{9i}{320}(\ga_a^{(1)} D_{(1)} \mu)-
\frac{21}{80}L_{(2)}(\ga_a^{(2)}\la)-
\frac{3}{40}L_{(4)}(\ga_a^{(4)}\la)\nn\\
&+\frac{3}{16}K_{(1)}(\ga_a^{(1)}\mu)
+\frac{63}{320}K_{(3)}(\ga_a^{(3)}\mu)
+\frac{9}{2560}(\la\ga_{(3)}\la)(\ga_a^{(3)}\mu)\nn\\
\widetilde{L}^{(2){}}_{}&=
-\frac{11}{150}L \mu -\frac{27}{200}L' \mu+\frac{7}{120}L_{(2)}(\ga^{(2)}\mu)
+\frac{1}{120}L_{(4)}(\ga^{(4)}\mu)\nn\\
&-\frac{7}{120}K_{(1)}(\ga^{(1)}\la)
-\frac{1}{24}K_{(3)}(\ga^{(3)}\la)
+\frac{1}{9600}(\mu\ga_{(3)}\mu)(\ga^{(3)}\la)~,
\end{align}
\vfill\break
\beal
\widetilde{L}^{(4)}_{abcd{}}&=\frac{1}{420}L_{(4)}(\ga_{abcd}^{(4)}\mu)\nn\\
\widetilde{L}^{(4)}_{abc{}}&=
-\frac{1}{168}L_{(4)}(\ga_{abc}^{(4)}\la)
-\frac{1}{336}K_{(3)}(\ga_{abc}^{(3)}\mu)
\nn\\
\widetilde{L}^{(4)}_{ab{}}&=-
\frac{1}{32}\widetilde{T}_{ab}+\frac{1}{432}L_{(2)}(\ga_{ab}^{(2)}\mu)
-\frac{13}{360}L_{(4)}(\ga_{ab}^{(4)}\mu)-\frac{1}{144}K_{(3)}(\ga_{ab}^{(3)}\la)\nn\\
\widetilde{L}^{(4)}_{a{}}&=
-\frac{i}{1920}(\ga_a^{(1)} D_{(1)} \mu)-
\frac{1}{480}L_{(2)}(\ga_a^{(2)}\la)+
\frac{31}{1680}L_{(4)}(\ga_a^{(4)}\la)\nn\\
&-\frac{1}{480}K_{(1)}(\ga_a^{(1)}\mu)
+\frac{11}{4480}K_{(3)}(\ga_a^{(3)}\mu)
+\frac{1}{15360}(\la\ga_{(3)}\la)(\ga_a^{(3)}\mu)\nn\\
\widetilde{L}^{(4)}_{{}}&=-\frac{1}{900}L \mu +\frac{11}{2400}L' \mu 
-\frac{1}{4320}L_{(2)}(\ga^{(2)}\mu)
-\frac{31}{10080}L_{(4)}(\ga^{(4)}\mu)
\nn\\
&-\frac{1}{1440}K_{(1)}(\ga^{(1)}\la)
-\frac{1}{1440}K_{(3)}(\ga^{(3)}\la)
+\frac{7}{115200}(\mu\ga_{(3)}\mu)(\ga^{(3)}\la)
\end{align}
and
\beal
\widetilde{K}^{(1)}_{a}&=
\frac{i}{20}(\ga_a^{(1)} D_{(1)} \mu)+\frac{1}{5}L_{(2)}(\ga_a^{(2)}\la)-
\frac{2}{5}L_{(4)}(\ga_a^{(4)}\la)\nn\\
&-\frac{3}{20}K_{(3)}(\ga_a^{(3)}\mu)
-\frac{1}{160}(\la\ga_{(3)}\la)(\ga_a^{(3)}\mu)\nn\\
\widetilde{K}^{(1)}&=-\frac{1}{25}L \mu+\frac{1}{25}L' \mu
 -\frac{2}{15}L_{(2)}(\ga^{(2)}\mu)
-\frac{2}{5}L_{(4)}(\ga^{(4)}\mu)\nn\\
&-\frac{3}{5}K_{(1)}(\ga^{(1)}\la)
-\frac{1}{5}K_{(3)}(\ga^{(3)}\la)
+\frac{1}{100}(\mu\ga_{(3)}\mu)(\ga^{(3)}\la)~,
\end{align}
\beal
\widetilde{K}^{(3)}_{abc}&=
-\frac{1}{84}K_{(3)}(\ga_{abc}^{(3)}\mu)\nn\\
\widetilde{K}^{(3)}_{ab}&=
\frac{1}{8}\widetilde{T}_{ab}+\frac{1}{108}L_{(2)}(\ga_{ab}^{(2)}\mu)
+\frac{1}{6}L_{(4)}(\ga_{ab}^{(4)}\mu)+\frac{1}{18}K_{(3)}(\ga_{ab}^{(3)}\la)\nn\\
\widetilde{K}^{(3)}_{a}&=
-\frac{i}{160}(\ga_a^{(1)} D_{(1)} \mu)-\frac{1}{40}L_{(2)}(\ga_a^{(2)}\la)+
\frac{1}{20}L_{(4)}(\ga_a^{(4)}\la)\nn\\
&+\frac{1}{40}K_{(1)}(\ga_a^{(1)}\mu)
+\frac{81}{1120}K_{(3)}(\ga_a^{(3)}\mu)
+\frac{1}{1280}(\la\ga_{(3)}\la)(\ga_a^{(3)}\mu)\nn\\
\widetilde{K}^{(3)}_{}&=\frac{13}{900}L \mu+\frac{1}{75}L' \mu
 +\frac{1}{540}L_{(2)}(\ga^{(2)}\mu)
+\frac{1}{180}L_{(4)}(\ga^{(4)}\mu)\nn\\
&+\frac{1}{90}K_{(1)}(\ga^{(1)}\la)
+\frac{1}{60}K_{(3)}(\ga^{(3)}\la)
-\frac{1}{7200}(\mu\ga_{(3)}\mu)(\ga^{(3)}\la)
~,
\end{align}
for the right-handed spinors $\widetilde{L}^{\al}$, 
$\widetilde{L}^{(2)\al}_{ab}$, etc. 
Similarly, for the left-handed spinors we have
\beal
\widetilde{L}^{{}}&=\widetilde{L}^{\prime}
+\frac{7}{2}L \la -2L' \la+\frac{3}{2}L_{(2)}(\ga^{(2)}\la)
+\frac{3}{2}L_{(4)}(\ga^{(4)}\la)\nn\\
&-\frac{3}{2}K_{(1)}(\ga^{(1)}\mu)
-\frac{3}{2}K_{(3)}(\ga^{(3)}\mu)
+\frac{1}{32}(\la\ga_{(3)}\la)(\ga^{(3)}\mu)~,
\label{leftspinorl}
\end{align}
\beal
\widetilde{L}^{(2){}}_{ab}&=
\frac{3}{16}\widetilde{T}_{ab}
-\frac{1}{24}L_{(2)}(\ga_{ab}^{(2)}\la)-\frac{1}{4}L_{(4)}(\ga_{ab}^{(4)}\la)
-\frac{1}{8}K_{(3)}(\ga_{ab}^{(3)}\mu)\nn\\
\widetilde{L}^{(2){}}_{a}&=
-\frac{9i}{320}(\ga_a^{(1)} D_{(1)} \la)-
\frac{21}{80}L_{(2)}(\ga_a^{(2)}\mu)+
\frac{3}{40}L_{(4)}(\ga_a^{(4)}\mu)\nn\\
&+\frac{3}{16}K_{(1)}(\ga_a^{(1)}\la)
-\frac{63}{320}K_{(3)}(\ga_a^{(3)}\la)
+\frac{9}{2560}(\mu\ga_{(3)}\mu)(\ga_a^{(3)}\la)\nn\\
\widetilde{L}^{(2){}}_{}&=
-\frac{27}{200}L \la -\frac{11}{150}L' \la +\frac{7}{120}L_{(2)}(\ga^{(2)}\la)
-\frac{1}{120}L_{(4)}(\ga^{(4)}\la)\nn\\
&-\frac{7}{120}K_{(1)}(\ga^{(1)}\mu)
+\frac{1}{24}K_{(3)}(\ga^{(3)}\mu)
+\frac{1}{9600}(\la\ga_{(3)}\la)(\ga^{(3)}\mu)~,
\end{align}
\beal
\widetilde{L}^{(4)}_{abcd{}}&=\frac{1}{420}L_{(4)}(\ga_{abcd}^{(4)}\la)\nn\\
\widetilde{L}^{(4)}_{abc{}}&=-\frac{1}{168}L_{(4)}(\ga_{abc}^{(4)}\mu)
-\frac{1}{336}K_{(3)}(\ga_{abc}^{(3)}\la)
\nn\\
\widetilde{L}^{(4)}_{ab{}}&=
\frac{1}{32}\widetilde{T}_{ab}
-\frac{1}{432}L_{(2)}(\ga_{ab}^{(2)}\la)
-\frac{13}{360}L_{(4)}(\ga_{ab}^{(4)}\la)
-\frac{1}{144}K_{(3)}(\ga_{ab}^{(3)}\mu)\nn\\
\widetilde{L}^{(4)}_{a{}}&=
\frac{i}{1920}(\ga_a^{(1)} D_{(1)} \la)+
\frac{1}{480}L_{(2)}(\ga_a^{(2)}\mu)+
\frac{31}{1680}L_{(4)}(\ga_a^{(4)}\mu)\nn\\
&+\frac{1}{480}K_{(1)}(\ga_a^{(1)}\la)
+\frac{11}{4480}K_{(3)}(\ga_a^{(3)}\la)
-\frac{1}{15360}(\mu\ga_{(3)}\mu)(\ga_a^{(3)}\la)\nn\\
\widetilde{L}^{(4)}_{{}}&=
-\frac{11}{2400}L \la +\frac{1}{900}L' \la +\frac{1}{4320}L_{(2)}(\ga^{(2)}\la)
-\frac{31}{10080}L_{(4)}(\ga^{(4)}\la)\nn\\
&+\frac{1}{1440}K_{(1)}(\ga^{(1)}\mu)
-\frac{1}{1440}K_{(3)}(\ga^{(3)}\mu)
-\frac{7}{115200}(\la\ga_{(3)}\la)(\ga^{(3)}\mu)
\end{align}
and
\beal
\widetilde{K}^{(1)}_{a}&=
\frac{i}{20}(\ga_a^{(1)} D_{(1)} \la)+\frac{1}{5}L_{(2)}(\ga_{a}^{(2)}\mu)
+\frac{2}{5}L_{(4)}(\ga_{a}^{(4)}\mu)\nn\\
&+\frac{3}{20}K_{(3)}(\ga_{a}^{(3)}\la)
-\frac{1}{160}(\mu\ga_{(3)}\mu)(\ga_{a}^{(3)}\la)
\nn\\
\widetilde{K}^{(1)}&=
\frac{1}{25}L \la -\frac{1}{25}L' \la -\frac{2}{15}L_{(2)}(\ga^{(2)}\la)
+\frac{2}{5}L_{(4)}(\ga^{(4)}\la)\nn\\
&-\frac{3}{5}K_{(1)}(\ga^{(1)}\mu)
+\frac{1}{5}K_{(3)}(\ga^{(3)}\mu)
+\frac{1}{100}(\la\ga_{(3)}\la)(\ga^{(3)}\mu)~,
\end{align}
\beal
\widetilde{K}^{(3)}_{abc}&=-\frac{1}{84}K_{(3)}(\ga_{abc}^{(3)}\la)\nn\\
\widetilde{K}^{(3)}_{ab}&=
-\frac{1}{8}\widetilde{T}_{ab}-
\frac{1}{108}L_{(2)}(\ga_{ab}^{(2)}\la)+
\frac{1}{6}L_{(4)}(\ga_{ab}^{(4)}\la)
+\frac{1}{18}K_{(3)}(\ga_{ab}^{(3)}\mu)\nn\\
\widetilde{K}^{(3)}_{a}&=
\frac{i}{160}(\ga_a^{(1)} D_{(1)} \la)+\frac{1}{40}L_{(2)}(\ga_{a}^{(2)}\mu)
+\frac{1}{20}L_{(4)}(\ga_{a}^{(4)}\mu)\nn\\
&-\frac{1}{40}K_{(1)}(\ga_{a}^{(1)}\la)+\frac{81}{1120}K_{(3)}(\ga_{a}^{(3)}\la)
-\frac{1}{1280}(\mu\ga_{(3)}\mu)(\ga_{a}^{(3)}\la)
\nn\\
\widetilde{K}^{(3)}_{}&=
-\frac{1}{75}L \la -\frac{13}{900}L' \la -\frac{1}{540}L_{(2)}(\ga^{(2)}\la)
+\frac{1}{180}L_{(4)}(\ga^{(4)}\la)\nn\\
&-\frac{1}{90}K_{(1)}(\ga^{(1)}\mu)
+\frac{1}{60}K_{(3)}(\ga^{(3)}\mu)
+\frac{1}{7200}(\la\ga_{(3)}\la)(\ga^{(3)}\mu)
~.
\end{align}

{\it Fermionic equations-of-motion}
\beal
\ga^bT_{ab}&=-4\widetilde{T}_a-9\ga_a\widetilde{T}~,
\end{align}
where $\widetilde{T}$, $\widetilde{T}_a$  are given in (\ref{mi}, \ref{ni}) and
\beal
i\slsh D \la&=-\frac{24}{5}(L-L') \mu-\frac{16}{3}L_{(2)}(\ga^{(2)}\mu)\nn\\
&-12K_{(1)}(\ga^{(1)}\la)
+3K_{(3)}(\ga^{(3)}\la)
+\frac{3}{40}(\mu\ga_{(3)}\mu)(\ga^{(3)}\la)\nn\\
i\slsh D \mu&=\frac{24}{5}(L-L') \la-\frac{16}{3}L_{(2)}(\ga^{(2)}\la)\nn\\
&-12K_{(1)}(\ga^{(1)}\mu)
-3K_{(3)}(\ga^{(3)}\mu)
+\frac{3}{40}(\la\ga_{(3)}\la)(\ga^{(3)}\mu)~.
\end{align}

\subsection*{Dimension ${2}$}

{\it Curvature}
\beal
R_{[abc]d}=R_{a[bcd]}=0~.
\end{align}

{\it Spinor derivatives}
\beal
D_{\una}T_{ab}{}^{\ub}=R_{ab\una}{}^{\ub}-2D_{[a}T_{b]\una}{}^{\ub}-2T_{\una[a|}{}^{\une}
T_{\une|b]}{}^{\ub}-T_{ab}{}^{\une}T_{\une\una}{}^{\ub}~.
\end{align}

{\it Bosonic equations-of-motion (HLW)}
\beal
R_{ab}&=18mD_{(a}A_{b)}+36m^2A_aA_b-\frac{1}{2}F_{a}{}^eF_{be}
-144H_{a}{}^{ef}H_{bef}\nn\\
&-3072G_{a}{}^{efg}G_{befg}
-\eta_{ab}\Big(36m^2+\frac{1}{4}F_{ef}F^{ef}-48H_{efg}H^{efg}\Big)\nn\\
md*A&=-18m^2A\wedge*A
+\frac{1}{4}F\wedge*F-96H\wedge*H
+3072G\wedge*G\nn\\
d*F&=-72m^2*A+18mA\wedge*F+4608H\wedge*G\nn\\
d*H&=-12m A\wedge* H+8F\wedge*G
+768G\wedge G\nn\\
d*G&=-\frac{3}{2}m*H+12mA\wedge*G+24H\wedge G~.
\label{hl1}
\end{align}

{\it Bosonic Bianchi identities (HLW)}
\beal
dA&= F\nn\\
dF&=0\nn\\
dH&=48mG-6mA\wedge H\nn\\
dG&=6mA\wedge G-\frac{1}{8}F\wedge H~,
\label{hl2}
\end{align}
where we have introduced the more conventional notation: 
$iK_{(1)}=\frac{3}{2 }mA$, 
$L_{(2)}=\frac{3}{16}F$, 
$iK_{(3)}=H$, 
$L_{(4)}=G$.  
This  is exactly the gauge-fixed form of the equations presented in \cite{hlw}\footnote{
To compare with the equations presented in \cite{hlw} one has to 
set the $\sigma$ field of that reference to zero. 
Remember that $\sigma$ is analogous to a St\"{u}ckelberg field and 
can be gauged-away for $m\neq 0$. 
Note also that there is a typographical error 
in the coefficient of the second term on the right-hand-side of the third equation  
of (4.3) of \cite{hlw}: instead of $1/4$ it should read $3/4$. This was subsequently 
corrected in \cite{cla}.}. 
HLW can also be obtained by a 
generalized Scherk-Schwarz reduction of ordinary 
eleven-dimensional supergravity \cite{llp} (see also \cite{ghee}). We have checked that 
the equations presented here indeed  
coincide with those in \cite{llp}\footnote{To bring the equations above 
in the form presented in that reference, one needs to substitute $m\rightarrow\frac{1}{2}m$, $F\rightarrow F_{(2)}$, 
$H\rightarrow -\frac{1}{24}F_{(3)}$, $G\rightarrow\frac{1}{192}F_{(4)}$ 
and set the pure-gauge field $\varphi$ of \cite{llp} to zero.}.

\subsection*{Dimension $\frac{5}{2}$}

{\it Spinor derivatives}

\beal
D_{\una}R_{abcd}
=2D_{[a|}R_{\una|b]cd}-T_{ab}{}^{\une} R_{\une \una cd}
+2T_{[a|\una}{}^{\une} R_{\une|b]cd} ~.
\end{align}

\section{Conclusions}
\label{conclusions}

In this paper we have employed a systematic procedure in order 
 to search for massive deformations of 
IIA supergravity. It is amusing to think that had 
 we not known about them, we would have been 
able to discover in this way both Romans and HLW supergravities in one go. 
The method used here is quite general; 
it would therefore be of interest to apply it to other 
supersymmetric systems. It is quite plausible that 
new massive supergravities can be discovered in this way.

As already mentioned in the introduction, HLW supergravity arises upon
compactification of a topologically  modified version of eleven-dimensional
supergravity, MM-theory. However, it is not known at present whether
MM-theory can be given a microscopic quantum-mechanical description. Given
that de Sitter space is an (unstable) vacuum of MM-theory, if the latter can
somehow be related to M-theory it would provide a mechanism for embedding de
Sitter in M-theory. This interesting direction deserves to be pursued
further, alongside with more recent proposals for the realization of de
Sitter space in string/M-theory \cite{kklt}.

In a step towards this direction, it was shown in \cite{clb} that HLW
supergravity supports (nonsupersymmetric) multi-zero-brane solutions. It was
also argued that these states may indeed be associated with a microscopic
description of MM-theory\footnote{In \cite{ramg} it was suggested that a
microscopic Matrix-model description of MM-theory may be obtainable by a
Euclidean radial reduction, as opposed to the usual dimensional reduction of
Matrix theory.} and that the latter should represent an unstable phase of
M-theory. A better understanding of the dynamics of these zero-branes is
important in testing the proposal of \cite{clb}. To that end it would be
interesting to construct 
the world-volume theories of 
`massive' kappa-symmetric objects propagating in a
HLW background (or perhaps directly in MM-theory)\footnote{
For the case of Romans supergravity, such `massive' branes 
were considered in \cite{berg, blo, hs}.}, 
either within the
superembedding formalism or by directly imposing kappa symmetry.

\appendix

\section{Gamma-traceless projections}
\label{gammatraceless}

In this appendix we explain our conventions concerning the projections onto the gamma-traceless part
of the various form-spinors which appear in the analysis of the BI's at dimension three-half.

Let $S$ be a spinor (it may be either chiral or antichiral) and $\Phi_{(p)}$ be a $p$-form. 
The following projections (used in section \ref{bi}) are gamma-traceless, as the reader may verify:

{\it Vector-spinor}
\beal
\Phi_{(1)}(\ga_a^{(1)}S)&:=\Phi_i (\ga^{i}{}_{a} S)  +9\Phi_{a} S \nn\\
\Phi_{(2)}(\ga_a^{(2)}S)&:=\Phi_{ij} (\ga^{ij}{}_{a} S)  +8\Phi_{ia}(\ga^{i} S) \nn\\
\Phi_{(3)}(\ga_a^{(3)}S)&:=\Phi_{ijk} (\ga^{ijk}{}_{a} S)  +7\Phi_{ija}(\ga^{ij} S) \nn\\
\Phi_{(4)}(\ga_a^{(4)}S)&:=\Phi_{ijkl} (\ga^{ijkl}{}_{a} S)  +6\Phi_{ijka}(\ga^{ijk} S)~.
\end{align}
{\it Two-form-spinor}
\beal
\Phi_{(2)}(\ga_{ab}^{(2)}S)&:= \Phi_{ij} (\ga^{ij}{}_{ab} S)  +14\Phi_{i[a}(\ga^{i}{}_{b]} S)  
-56\Phi_{ab} S  \nn\\
\Phi_{(3)}(\ga_{ab}^{(3)}S)&:=\Phi_{ijk} (\ga^{ijk}{}_{ab} S)  +12\Phi_{ij[a}(\ga^{ij}{}_{b]} S)  
-42\Phi_{iab}(\ga^{i} S)  \nn\\
\Phi_{(4)}(\ga_{ab}^{(4)}S)&:=\Phi_{ijkl} (\ga^{ijkl}{}_{ab} S)  +10\Phi_{ijk[a}(\ga^{ijk}{}_{b]} S)  
-30\Phi_{ijab}(\ga^{ij} S)~.
\end{align}
{\it Three-form-spinor}
\beal
\Phi_{(3)}(\ga_{abc}^{(3)}S)&:=\Phi_{ijk} (\ga^{ijk}{}_{abc} S)  +15\Phi_{ij[a}(\ga^{ij}{}_{bc]} S)  
-90\Phi_{i[ab}(\ga^{i}{}_{c]} S)-210\Phi_{abc}S \nn\\
\Phi_{(4)}(\ga_{abc}^{(4)}S)&:=\Phi_{ijkl} (\ga^{ijkl}{}_{abc} S)  +12\Phi_{ijk[a}(\ga^{ij}{}_{bc]} S)  
-60\Phi_{ij[ab}(\ga^{ij}{}_{c]} S)-120\Phi_{iabc}(\ga^iS) ~.
\end{align}
{\it Four-form-spinor}
\beal
\Phi_{(4)}(\ga_{abcd}^{(4)}S):=\Phi_{ijkl} (\ga^{ijkl}{}_{abcd} S)  +&12\Phi_{ijk[a}(\ga^{ijk}{}_{bcd]} S)  
-72\Phi_{ij[ab}(\ga^{ij}{}_{cd]} S)\nn\\
-&240\Phi_{i[abc}(\ga^{i}{}_{d]} S)+360\Phi_{abcd}S ~.
\end{align}

\vfill\break

%
%

\end{document}